\documentclass[journal]{IEEEtran}
\ifCLASSINFOpdf
\else
\fi
\usepackage{cite}
\usepackage{amsmath,amssymb,amsfonts}
\usepackage{algorithmic}
\usepackage[backref]{hyperref} 
\hypersetup{
hidelinks
}
\usepackage{graphicx}
\usepackage{url}
\usepackage{algorithm}
\usepackage{textcomp}
\usepackage[caption=false,farskip=0pt,labelfont={bf}]{subfig}
\usepackage[T1]{fontenc}
\usepackage{booktabs} 
\usepackage{xcolor}
\usepackage{multirow}
\usepackage{color}
\usepackage{newclude}

\begin{document}
\begin{sloppypar}
%
\title{LUK: Empowering Log Understanding with Expert Knowledge from Large Language Models}
%
%
%


\author{Lipeng~Ma,
        Weidong~Yang,
        Sihang~Jiang,
        Ben~Fei,
        Mingjie~Zhou,
        Shuhao~Li,
        Mingyu~Zhao,
        Bo~Xu,
        and~Yanghua~Xiao
    
\thanks{This work was supported by the National Natural Science Foundation of
China (U2033209) and the Natural Science Foundation of Shanghai (No. 24ZR1405000). \textit{Corresponding authors: Weidong Yang and Bo Xu}}
\thanks{Lipeng Ma, Weidong Yang, Sihang Jiang, Ben Fei, Mingjie Zhou, Shuhao Li and Yanghua Xiao are with the School of Computer Science, Fudan University, Shanghai, China, 200433 (e-mails: $\{\text{lpma21}, \text{bfei21}, \text{shli23}\}$@m.fudan.edu.cn;  $\{\text{wdyang},\text{jiangsihang},\text{mjzhou19},\text{shawyh}\}$@fudan.edu.cn).}
\thanks{Mingyu Zhao is with the School of Artificial Intelligence, Optics and Electronics (iOPEN), Northwestern Polytechnical University, Xi'an, Shaanxi, China, 710072. (email: myzhao@nwpu.edu.cn)}
\thanks{Bo Xu is with the School of Computer Science and Technology, Donghua University, Shanghai, China, 201620 (e-mails: xubo@dhu.edu.cn)}
}

%
%


\markboth{Journal of \LaTeX\ Class Files,~Vol.~14, No.~8, August~2021}%
{Shell \MakeLowercase{\textit{et al.}}: A Sample Article Using IEEEtran.cls for IEEE Journals}

%



\maketitle

\begin{abstract}
Logs play a critical role in providing essential information for system monitoring and troubleshooting. 
Recently, with the success of pre-trained language models (PLMs) and large language models (LLMs) in natural language processing (NLP), smaller PLMs (such as BERT) and LLMs (like GPT-4) have become the current mainstream approaches for log analysis.
\textcolor{black}{
Despite the remarkable capabilities of LLMs, their higher cost and inefficient inference present significant challenges in leveraging the full potential of LLMs to analyze logs.}
In contrast, smaller PLMs can be fine-tuned for specific tasks even with limited computational resources, making them more practical. However, these smaller PLMs face challenges in understanding logs comprehensively due to their limited expert knowledge.
\textcolor{black}{To address the lack of expert knowledge and enhance log understanding for smaller PLMs, this paper introduces a novel and practical knowledge enhancement framework, called LUK, which acquires expert knowledge from LLMs automatically and then enhances the smaller PLM for log analysis with these expert knowledge. LUK can take full advantage of both types of models.} 
Specifically, we design a multi-expert collaboration framework based on LLMs with different roles to acquire expert knowledge. In addition, we propose two novel pre-training tasks to enhance the log pre-training with expert knowledge. LUK achieves state-of-the-art results on different log analysis tasks and extensive experiments demonstrate expert knowledge from LLMs can be utilized more effectively to understand logs.
Our source code and detailed experimental data are available at \url{https://github.com/LeaperOvO/LUK}.

\end{abstract}


%
\IEEEpeerreviewmaketitle
\section{Introduction}
\IEEEPARstart{W}{ith} the increasing complexity and scale of IT systems,  the maintenance and operation of large-scale systems become more challenging. Logs record the valuable runtime status, they play a crucial role in troubleshooting and enable engineers to monitor system health effectively. However, the increasing volume of logs has made manual analysis a harder task \cite{zhang2021onion}. Consequently, numerous automated log analysis methods utilizing machine learning (ML) or deep learning (DL) models have been proposed to automatically analyze logs, encompassing various tasks such as log parsing \cite{zhu2019tools,dai2020logram,nedelkoski2021self,liu2022uniparser}, anomaly detection \cite{du2017deeplog,meng2019loganomaly,li2020swisslog,nedelkoski2020self,huang2020hitanomaly,han2021unsupervised,le2022log}, root cause analysis \cite{lu2017log,wang2020root,soldani2022anomaly}, failure prediction \cite{zhang2018prefix,gao2020task,cotroneo2019bad}, etc. In particular, with the recent success of pre-trained language models (PLMs) in natural language processing (NLP), especially the advent of large language models (LLMs) represented by ChatGPT and GPT-4 \cite{achiam2023gpt},  language models (LMs) have garnered significant attention in log understanding and these LM-based approaches \cite{zhu2021unilog,tao2023biglog,ma2024knowlog, le2023log, lee2023lanobert,almodovar2024logfit, xu2024divlog,xu2024unilog,liu2023logprompt,jiang2023llmparser,guo2024lemur} achieve tremendous achievements in automated log analysis due to their outstanding performance. 





\begin{figure}[]
\centering
   \includegraphics[width=\linewidth]{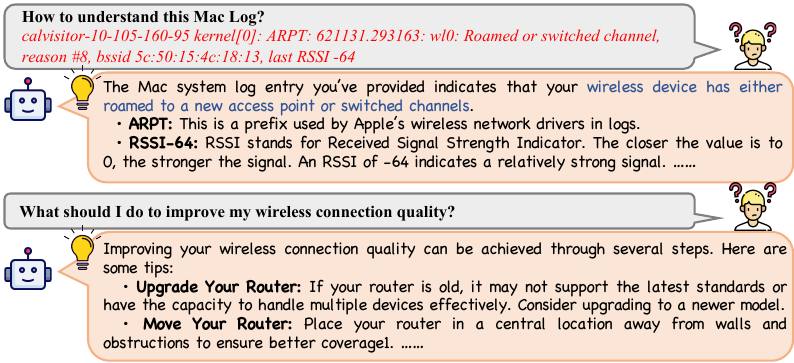}
   \vspace{-3mm}
   \caption{Example of prompting ChatGPT to help understand log, which can provide detailed background information and terminology explanations and solutions. Without knowledge, the beginner fails to understand RSSI value.}
   \vspace{-3mm}
   \label{fig:understand}
\end{figure}

As language models increase in scale and gain enhanced capabilities, two mainstream paradigms have emerged for utilizing LMs in log understanding. The first one is based on PLMs, which follows the \textit{pre-train \& fine-tune} paradigm, such as BERT \cite{devlin2019bert}, fine-tuning a PLM with task-specific data and enabling it to specialize in the given task. 
Considering the need for fine-tuning with limited computational resources and cost savings, LMs in this paradigm are typically smaller in size, such as BERT with 110M parameters.
The second one is based on LLMs with vast scale and complexity\footnote{We use LLM to specifically refer to the large language models over 10B parameters that utilize the ICL paradigm without fine-tuning the parameters.}, which can tackle a diverse array of tasks with remarkable proficiency. The LLM-based methods follow the \textit{In-Context Learning} (ICL) paradigm \cite{dong2022survey,lu2022fantastically}, learning from a few examples in the context without updating the parameters. 

\textcolor{black}{
Despite the powerful performance of LLMs, limited accessibility and higher costs present significant challenges in leveraging the full potential of LLMs. 
These LLMs often come with substantial
usage fees and have restricted access to data privacy and security. Even if the LLM is deployed locally, it still may face resource constraints, making LLMs
less attainable for individuals and smaller organizations.  
In addition, due to their
complex autoregressive architectures and a large number of parameters, LLM-based methods are computationally
inefficient.
Considering that modern software systems can produce several petabytes per day \cite{wang2022spine,li2024logshrink}, directly employing LLMs for log analysis is impractical due to the significant overhead of querying LLMs, such as inference time and network latency \cite{jiang2023llmparser,wang2023tabi}. 
In contrast, smaller PLMs with smaller parameters, requiring limited resource costs, are more efficient than LLMs \cite{ma2024llmparser}. 
Hence, we argue that the smaller PLM is more practical in extensive log analysis scenarios \cite{le2023log,lee2023lanobert,almodovar2024logfit}.
}

However, the smaller PLM encounters a bottleneck in log understanding due to the inadequacy of expert knowledge. 
Logs inherently employ concise and highly specialized terminology, 
the smaller PLM struggles to understand valuable information from logs like experts \cite{ma2024knowlog}. 
\textcolor{black}{This issue can be attributed to the lack of rich sources for knowledge acquisition. Although KnowLog proposes to acquire expert knowledge from documentation \cite{ma2024knowlog}, most logs lack professional documentation (e.g. almost unable to find relevant documents for logs coming from Loghub \cite{zhu2023loghub}) and the manual construction of knowledge is costly and inefficient, which significantly hinders the availability of the model.
}


Fortunately, LLMs with extensive knowledge can understand logs comprehensively, attributing to training on massive textual data related to code \cite{peng2023generative} and logging \cite{mastropaolo2022using,li2023exploring}. For example, as depicted in Fig. \ref{fig:understand}, ChatGPT can better assist in understanding the Mac log. \textcolor{black}{Consequently, LLMs such as GPT-4 compensate for the lack of knowledge of the smaller PLM, which can analyze logs more efficiently and professionally.}

To more effectively harness the knowledge embedded within LLMs for log understanding, we argue that expert knowledge can be elicited from LLMs and subsequently utilized to empower the smaller PLM, making the smaller model understand logs like experts.  However, two main challenges exist in enhancing log understanding with expert knowledge from LLMs. 
(1) \textbf{Effective Knowledge Acquisition:} 
\textcolor{black}{Although LLMs have shown remarkable capabilities in various tasks, LLMs still have difficulty in acquiring complete and accurate knowledge due to hallucination, which may result in not covering all the depth of knowledge and there may be misunderstandings when dealing with domain-specific data \cite{mundler2023self,manakul2023selfcheckgpt}. 
In addition, in the era of LLMs, knowledge distillation (KD) \cite{xu2024survey} emerges as a pivotal methodology for transferring advanced capabilities from LLMs to smaller models. 
Compared to obtaining answers directly from the teacher's model, distillation towards the reasoning process has proved to be a more effective method \cite{xu2024survey,zhu2024distilling}.
However, current knowledge distillation methods, such as COT-based KD \cite{hsieh2023distilling,shridhar2023distilling}, struggle to acquire expert knowledge, instead imitating the teacher’s reasoning forms and ignoring the premise that reasoning requires expert knowledge. These distillation methods, which emphasize the process of reasoning and outcome, still hardly compensate for the lack of expert knowledge.}
(2) \textbf{Knowledge Enhancement:} 
\textcolor{black}{Since logs and expert knowledge obtained from LLM are heterogeneous data, smaller models with fewer parameters and limited learning capacity struggle to capture knowledge \cite{wei2022emergent,lu2023emergent}. 
Masked Language Modeling (MLM) \cite{devlin2019bert} as the widely used pre-training task for most pre-trained models, they can only perceive their own contextual information and cannot incorporate external knowledge. In addition, KnowLog proposes an abbreviation prediction task specialized for log understanding. However, its ability only to enhance the knowledge of abbreviations in the logs is limited. It cannot adaptively perceive other important information from the knowledge to understand logs, such as the parameter knowledge of logs.
}

To overcome the above-mentioned challenges, we propose a novel knowledge enhancement framework to empower log understanding on a smaller PLM, called \textbf{LUK}. Rather than utilizing LLMs to solve specific tasks directly, LUK first acquires expert knowledge from LLMs, then enhances the log pre-training with the corresponding expert knowledge, and finally the knowledge-enhanced PLM for logs can be fine-tuned to solve downstream log analysis tasks efficiently.
\textcolor{black}{This framework can maximize the advantages of both LLM and smaller PLMs.}

Specifically, to solve the first issue, we design a multi-expert collaboration framework to acquire expert knowledge from LLMs. Motivated by the waterfall model \cite{petersen2009waterfall} in software engineering, we build a professional team consisting of \textit{Director}, \textit{Executor}, and \textit{Evaluator}, and define the identity and responsibility of the roles through prompt, enabling LLMs to think and handle tasks just like role play. 
Then the team cooperates and interacts to construct expert knowledge.
To solve the second issue, we propose two novel pre-training tasks with knowledge enhancement to incorporate and perceive knowledge into the pre-trained model: word-level token prediction and sentence-level semantic alignment.


To evaluate the effectiveness of LUK, we conduct experiments on the software system and network device logs, including six log analysis downstream tasks.
The experiment results show that LUK can harness knowledge from LLMs more effectively to improve log understanding and achieve state-of-the-art results on different log analysis tasks. In addition, LUK exhibits remarkable generalization and robustness in log analysis, with notable advantages in low-resource scenarios.

Our main contributions can be summarized as follows:
\begin{itemize}
    \item We propose a novel knowledge enhancement framework dubbed LUK, which leverages expert knowledge from LLMs to empower log understanding on a smaller PLM. To the best of our knowledge, we are the first to elicit expert knowledge of logs from LLMs to empower log understanding, offering a new perspective on log analysis.
    \item We design a multi-expert collaboration framework to acquire expert knowledge from LLMs automatically. Moreover, we propose two novel pre-training tasks to enhance the log pre-training on a smaller PLM with knowledge, which requires less computation costs to predict than solving specific tasks with LLM directly.
    \item LUK achieves state-of-the-art results on different log understanding tasks proving the effectiveness of acquiring knowledge from LLMs to empower log understanding. Moreover, LUK demonstrates significant generalization and robustness in log analysis, particularly showcasing distinct advantages in low-resource scenarios.
\end{itemize}

 

\section{Related Work and Motivation}

\subsection{\textcolor{black}{Pre-trained Language Models for Log Analysis}}
Pre-training is the key to the success of the PLMs and LLMs, where a language model is first trained on the extensive corpus with effective pre-training tasks to capture general knowledge and then be adapted to solve different downstream tasks \cite{bai2021pre,niu2022spt}. 
The emergence of BERT \cite{devlin2019bert} heralds the success of the \textit{pre-train \& fine-tune} paradigm in natural language. Such language models improve the model's ability on a specific task by fine-tuning using task-specific objective functions after pre-training. 
Many studies \cite{gu2021domain,beltagy2019scibert,gururangan2020don} have demonstrated that pre-training on domain corpus and fine-tuning with supervised data can improve performance on a specific task, even outperforming LLMs \cite{yang2023supervised,juneja2023small}.
In the field of log analysis, many studies including pre-training on log corpus \cite{zhu2021unilog,tao2023biglog,ma2024knowlog,guo2021logbert} and fine-tuning on log analysis tasks \cite{lee2023lanobert,almodovar2024logfit} have demonstrated the effectiveness of the pre-trained model for log analysis. 

However, training solely on general or log corpus struggles to further improve log understanding due to the lack of expert knowledge. KnowLog \cite{ma2024knowlog} firstly proposes to enhance log understanding by integrating expert knowledge during the pre-training phase. Nevertheless, KnowLog's usage scenarios are limited as it heavily relies on documentation created by human experts to acquire knowledge. In contrast, this work addresses more generalized log analysis scenarios where logs are available without accompanying expert knowledge.
\textcolor{black}{
In addition, these pre-trained models rely on the existing Masked Language Modeling (MLM) as the pre-training task, they only perceive their own contextual information and are unable to integrate external knowledge.
Although KnowLog also proposes the abbreviation prediction task, this method cannot adaptively perceive other knowledge than the abbreviation. Considering the abundance of knowledge required for log understanding, hence, we also investigate making the PLM adaptive in perceiving useful information from the knowledge during pre-training.}


\subsection{\textcolor{black}{Large Language Model for Log Analysis}}
\textcolor{black}{Large language models (LLMs), representing a remarkable advancement in artificial intelligence, have been trained on diverse corpora, enabling them to generate human-like text and provide accurate responses to queries \cite{kasneci2023ChatGPT}. Notably, the recent introduction of transformative models like ChatGPT and GPT-4, characterized by their enormous scale and alignment with human feedback, presents a novel opportunity for enhancing software engineering tasks \cite{white2023prompt,le2023log,feng2024prompting,wang2024software}.
}
LLMs can learn from the prompt context without training the model, which is called In-context Learning (ICL) \cite{brown2020language}. Furthermore, research has shown that augmenting the size of LLMs significantly boosts their capacity for knowledge encoding and reasoning \cite{wei2022chain, wei2022emergent}.
Due to the outstanding abilities of LLMs, many recent studies \cite{xu2024divlog,xu2024unilog,liu2023logprompt,jiang2023llmparser,guo2024lemur} utilize the ICL paradigm of LLMs for different log analysis tasks without fine-tuning the model and make tremendous achievements in log analysis tasks. 

\textcolor{black}{However, restricted accessibility, higher costs, and computational inefficiency limit the widespread use of LLM in real-world scenarios. For example, the GPT-3 model with 175 billion parameters requires 326GB of GPU memory to deploy \cite{frantar2022optq}. Given the enormous data volumes produced by modern software systems daily, the overhead of querying LLMs, including inference time and network latency, renders direct application of LLMs for log analysis impractical. In contrast, small PLMs with smaller parameters are more efficient compared to LLMs \cite{ma2024llmparser}.} We argue that the smaller PLM can make predictions better in log analysis tasks given sufficient expert knowledge. In this paper, we guide LLMs as domain experts with prompts and explore how to effectively acquire and utilize expert knowledge from LLMs to empower log understanding.

\begin{figure}[]
\centering
   \includegraphics[width=\linewidth]{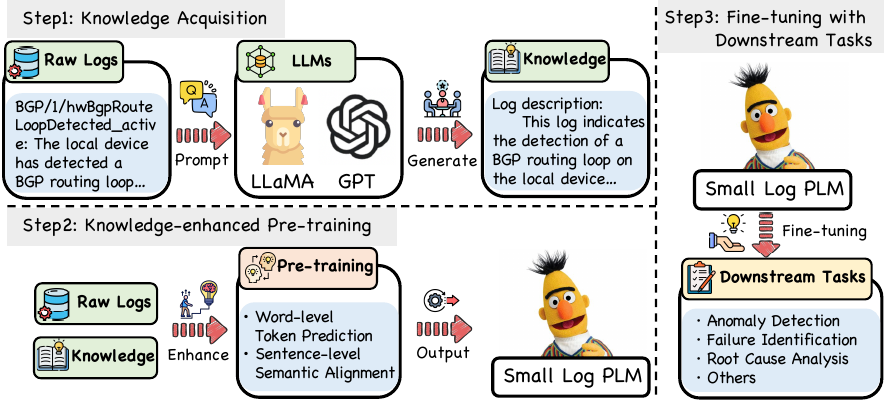}
   \vspace{-3mm}
   \caption{Conceptual overview of LUK.}
   \vspace{-3mm}
   \label{fig:overview}
\end{figure}

\section{Method}

\subsection{Overview}
Fig. \ref{fig:overview} shows the conceptual overview of LUK, which consists of three phases: knowledge acquisition, knowledge-enhanced pre-training, and fine-tuning with downstream tasks.
Specifically, in the first stage, we collect various logs as input, and then we design a multi-expert collaboration framework based on LLMs to acquire expert knowledge of logs.
In the second stage, we take the raw logs and the corresponding expert knowledge as input. To effectively leverage expert knowledge to empower log understanding on a smaller model, we enhance the log pre-training with knowledge based on BERT-base and propose two pre-training tasks on word level and sentence level.
Finally, a specific PLM for logs is obtained, which can be fine-tuned on downstream tasks to solve log analysis tasks.  
In the following sections, we describe the details of LUK.

\begin{figure*}[]
\centering
   \includegraphics[width=\textwidth]{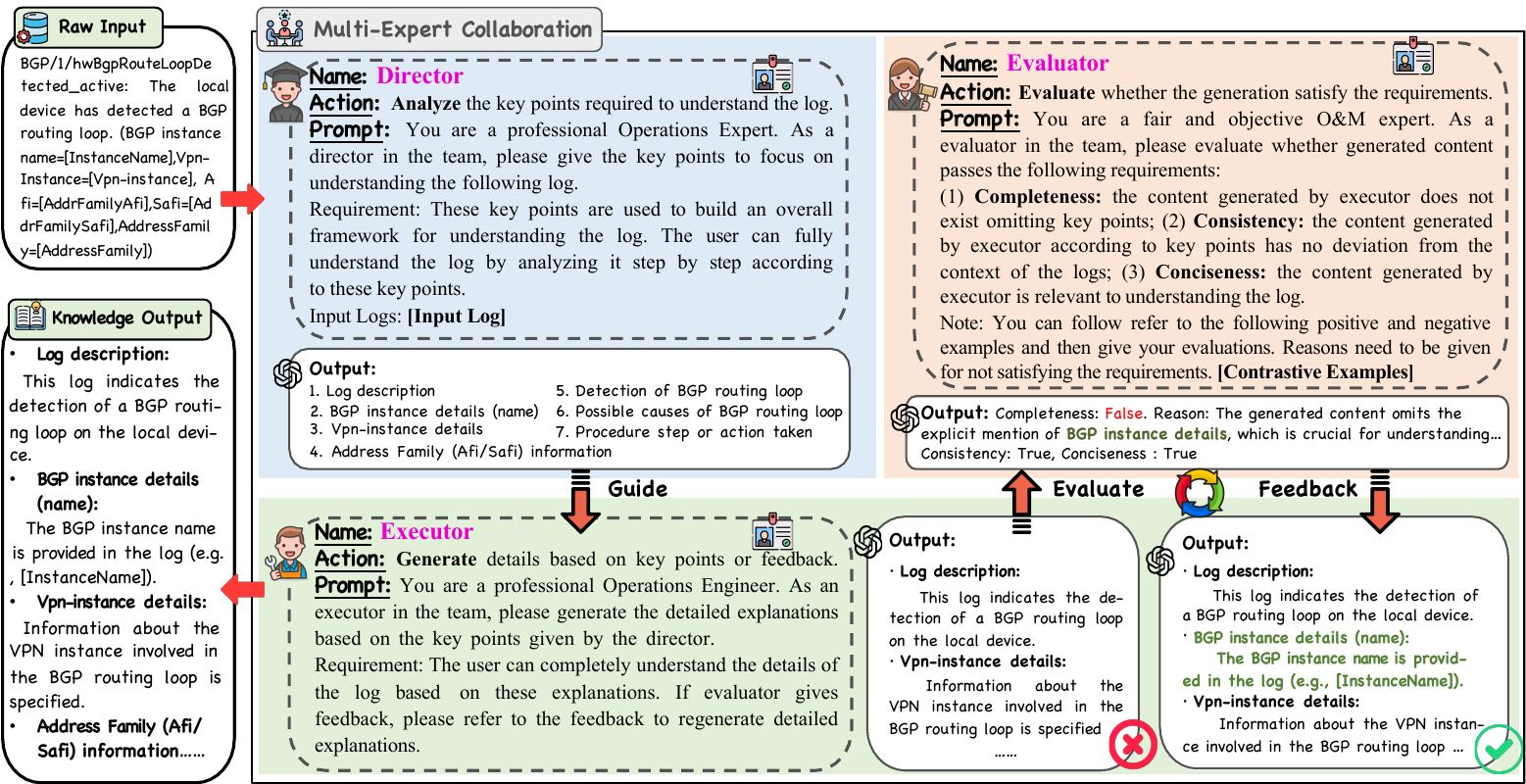}
   \vspace{-3mm}
   \caption{\textcolor{black}{Framework of multi-expert collaboration, which consists of three experts. First, the \textit{Director} meticulously analyzes the input logs to discern key points. Subsequently, the \textit{Executor} utilizes these key points
to create detailed content. Finally, the \textit{Evaluator} evaluates the quality of the generated content. In cases where the evaluation result is unsatisfactory and improvements are needed, the \textit{Evaluator} provides feedback to the \textit{Executor} for refinement. }}
   \vspace{-3mm}
   \label{fig:multi-expert}
\end{figure*}

\subsection{Multi-Expert Collaboration Framework}
\textcolor{black}{Hallucination is an inherent flaw of LLMs, which may lead to incomplete or incorrect generation, ultimately failing to fulfill
the intended requirements. To construct comprehensive and accurate expert knowledge with LLM,} motivated by \textit{cognitive synergy} \cite{goertzel2009cognitive} and \textit{teamwork theory} \cite{katzenbach2015wisdom}, humans can leverage the power of collaboration and interaction to solve complex problems. As shown in Fig. \ref{fig:multi-expert},  we design a multi-expert collaboration (MEC) framework, where we assign LLMs to different roles, and these roles collaborate to generate effective expert knowledge.

\subsubsection{Role Design}
Based on the classical waterfall model \cite{petersen2009waterfall} in software engineering, we design a similar waterfall model to analyze logs consisting of three stages: analysis, execution, and evaluation. Thus, we build a professional team based on LLMs and define clear goals for each role, comprising a \textit{Director}, \textit{Executor} and \textit{Evaluator}. 

\textcolor{black}{
It is widely acknowledged that one key capability of LLMs is to follow input natural language instructions or prompts \cite{zhou2023instruction,li2024evaluating}, hence they can customize the output to the specific needs of the user.
As shown in Fig. \ref{fig:multi-expert}, the prompt design methodology involves the pre-definition of role cards to allocate distinct roles and responsibilities to LLMs. These role cards encompass character identity, task objectives, requirements, and input queries. Initially, the LLM is assigned a character identity to emulate a specific expert role. This step enables the model to adjust its output to match the expertise and responsibilities associated with the designated role. Subsequently, the task objective is delineated to direct the LLM towards fulfilling the corresponding responsibilities aligned with the assigned character identity. Moreover, the requirement outlines the anticipated outcome or characteristics that the generated content must exhibit. These criteria may encompass aspects such as reasoning strategies, informativeness, or other pertinent factors crucial to the task at hand. Lastly, the inclusion of an input query or data serves to contextualize the task for the LLM by presenting specific information or scenarios that the model should engage with or respond to.
}

\textcolor{black}{
By integrating these components into the prompt design framework, we can effectively guide the LLM in generating responses that are tailored to specific roles, tasks, and desired outcomes.}
These three different roles are assigned the following tasks: 
\begin{itemize}
    \item \textbf{\textit{\textcolor{black}{Director}}.} 
    \textcolor{black}{The \textit{Director}'s primary objective is to establish a comprehensive high-level framework aimed at providing a strategic overview to drive the team's efforts in understanding logs. This role outlines the key points essential for log understanding, thus ensuring that essential insights are highlighted and comprehensively addressed.}
    \item \textbf{\textit{\textcolor{black}{Executor}}.} \textcolor{black}{As the central role of this team, the \textit{Executor} is tasked with the detailed execution of the log analysis process. By receiving key points from the Director and feedback from the Evaluator, the Executor undertakes content generation and refinement two critical responsibilities.  Firstly, the Executor generates content intricately aligned with the outlined key points. Secondly, the Executor refines and polishes the content based on feedback, enhancing the knowledge quality.}
    \item \textbf{\textit{\textcolor{black}{Evaluator}}.} \textcolor{black}{The Evaluator assumes the crucial responsibility of ensuring the quality and adherence of generated content to predefined requirements. 
    We define three evaluation requirements for the \textit{Evaluator} to scrutinize the Executor's generation on three fundamental aspects: completeness, consistency, and conciseness. By evaluating content against these criteria, the Evaluator validates the accuracy and completeness of the generated knowledge.}

\end{itemize}


\subsubsection{\textcolor{black}{Dehallucination with Self-Evaluation}}
\textcolor{black}{LLM Hallucinations seriously affect the accuracy and quality of outputs \cite{dhuliawala2023chain,huang2023survey}. Given the propensity of LLM-generated content to exhibit deficiencies in comprehensiveness and informativeness, ensuring the superior quality of directly generated knowledge poses a significant challenge.
To mitigate the hallucination of LLM and ensure the quality of expert knowledge, we build the role of an \textit{Evaluator} in the MEC framework and define three evaluation aspects to evaluate the quality of generated knowledge automatically,
which encourages the LLM to actively seek more detailed suggestions from the self-evaluation before delivering a formal response.
However, the MEC framework runs under the challenging reference-free setting and the evaluation requirements cannot be directly quantified through metrics, the LLM tends to generate ambiguous and inconsistent evaluations without fine-grained distinguishability, causing unsatisfactory evaluation performance \cite{zheng2023judging,ke2024critiquellm}.
}

\textcolor{black}{
To address the evaluation weaknesses of the LLM, we provide evaluation examples as the reference in the \textit{Evaluator} prompt. 
By providing specific examples, the \textit{Evaluator} uses them as the basis for evaluation, thus better steering the evaluation of LLMs towards our desired goals.
Considering that the single type of example makes the LLM biased, motivated by contrastive learning techniques \cite{gao2020dialogue,radford2021learning}, we construct \textbf{contrastive examples} consisting of positive and negative knowledge of the same log to \textit{Evaluator} as reference examples, where the positive example showcases the desired knowledge, while the negative example highlights the unsatisfactory knowledge that should be avoided.  
By analyzing both types of examples before generating the evaluation, the \textit{Evaluator} can reason about our evaluation requirements and make a more informed decision about how to evaluate. }

\begin{figure}[]
\centering
   \includegraphics[width=\linewidth]{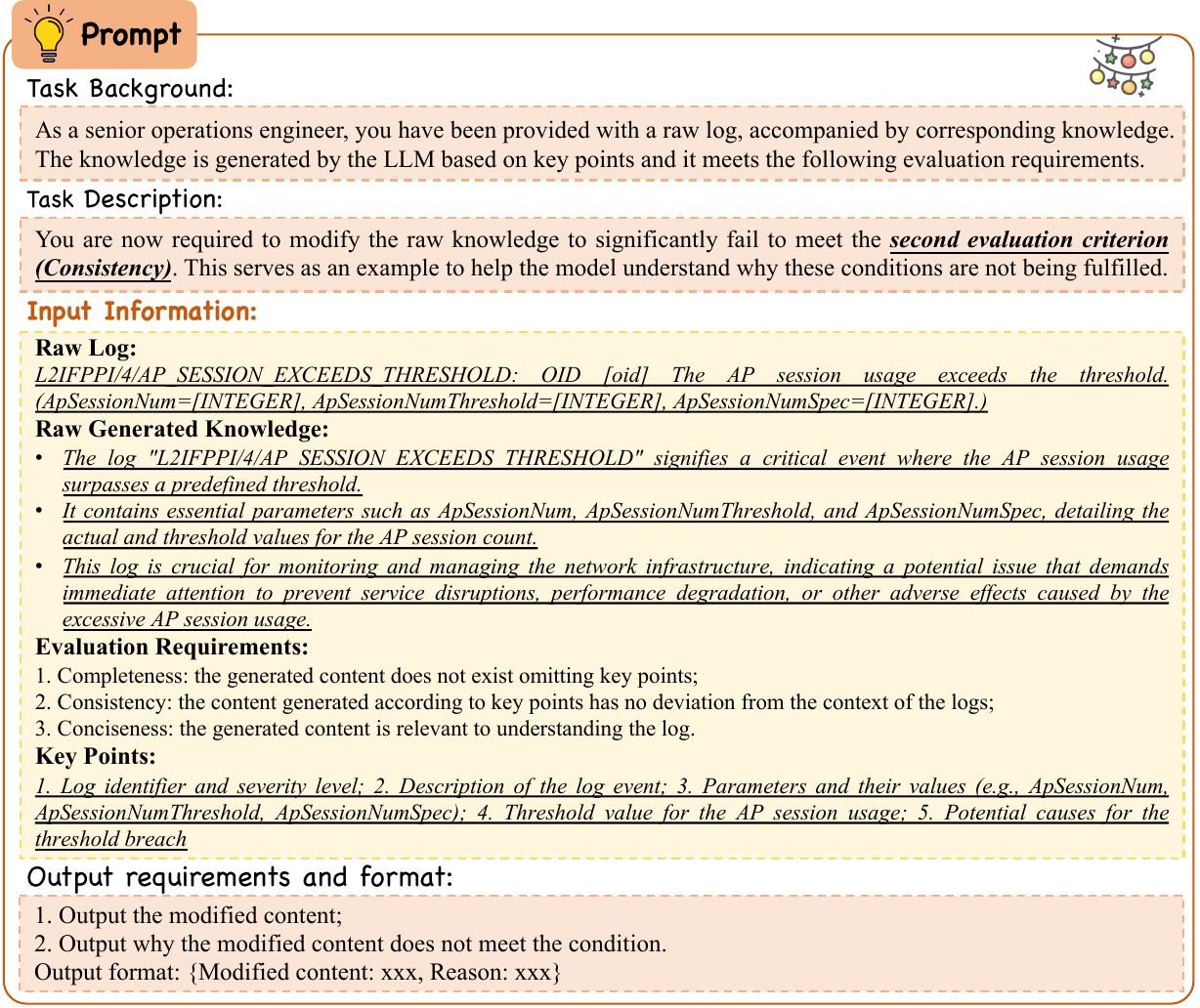}
   \caption{\textcolor{black}{Prompt template of constructing negative samples, where the underlined portion of the prompt supports user modification based on their own data and intent.}}
   \label{fig:negative_template}
\end{figure}

\textcolor{black}{
Specifically, we select one log $L$ with high-quality knowledge $K_{pos}$ as the positive example, which can be manually constructed by experts or generated from the LLM and manually confirmed. Considering that negative examples $\{K_{neg}\}_{c=1} ^ C, C=3$ of the log $L$ are unavailable directly, to construct the negative examples automatically, we propose to prompt the LLM to modify the positive knowledge making it unsatisfactory for evaluation.
As shown in Fig. \ref{fig:negative_template}, we first give the background in the prompt and then clarify the task of modifying knowledge for the LLM, where we can specify which evaluation metrics are not fulfilled. Next, we provide the raw log, the raw positive knowledge, the key points, and the three evaluation requirements in the prompt.
Finally, we require the LLM to give the reason for not fulfilling the evaluation requirements while generating the modified content, which not only improves the reasoning ability of the \textit{Evaluator} but also provides guidance for the revision of the content by \textit{Executor}.
In cases where the LLM-generated negative examples do not capture the characteristics we want to avoid, this can be manually checked, and feedback can be given to the LLM to re-generate negative samples. 
To avoid the higher expense of the longer prompt, we construct one positive knowledge and its corresponding three negative knowledge as reference examples to be provided to the \textit{Evaluator}. 
The contrastive examples we constructed in our experiments are in Fig. \ref{fig:evaluation}.
}

\subsubsection{Collaboration}
\textcolor{black}{After constructing evaluation examples and assigning roles to LLMs, different roles start working to collaborate and interact according to the requirements in the prompt. 
First, the \textit{Director} meticulously analyzes the input logs to discern key points and formulates a plan for comprehending the logs effectively. Subsequently, the \textit{Director} shares these insights with the \textit{Executor} to guide the content generation process. The \textit{Executor} utilizes these key points to create detailed content that aligns with the requirements.
Finally, the detailed content produced by the \textit{Executor} undergoes evaluation by the \textit{Evaluator} to assess its quality. If the content meets the evaluation requirement, it is directly output as expert knowledge. In cases where the evaluation result is unsatisfactory and improvements are needed, the \textit{Evaluator} provides feedback to the \textit{Executor} for refinement.  To rigorously ensure the quality of the generated knowledge, this evaluation and refinement mechanism will be iterative until the content generated by \textit{Executor} meets the evaluation requirement or the maximum number of iterations is reached. We set up 3 iterations to build expert knowledge in our experiments. The pseudocode of the multi-expert collaboration framework is outlined in Algorithm \ref{alg:algorithm}.
}

\textcolor{black}{ This collaborative framework manifests two pivotal advantages:
Firstly, leveraging multiple roles' collective insights and skills enables a multi-perspective analysis of the input log, 
thereby mitigating the one-sidedness and bias inherent in a single model's response acquiring more professional expert knowledge. 
Secondly, through the feedback loop mechanism, the framework enables iterative refinement of content. This iterative process of evaluation, feedback incorporation, and content enhancement promotes continuous improvement, thereby ensuring that the expert knowledge acquired is comprehensive and of high quality.}

\begin{algorithm}
\caption{\textcolor{black}{Pseudocode of multi-expert collaboration.}}
\label{alg:algorithm}
\begin{algorithmic}
\REQUIRE \textcolor{black}{Input Log $log$, LLM $\mathcal{M}$, Example Knowledge   $K_{pos}$}      
\ENSURE \textcolor{black}{Output Expert Knowledge $k$}
\STATE \textcolor{gray}{\# Construct Contrastive Examples}
\FOR{\textcolor{black}{$c = 1$ to $C$}}
    \STATE \textcolor{black}{$K_{neg}^c = \mathcal{M}(Prompt_{Modify}, K_{pos})$}
\ENDFOR
\STATE  \textcolor{black}{$K_{examples} = K_{pos} \cup \{K_{neg}\}_{c=1} ^ C $}
\STATE \textcolor{gray}{\# Initial Roles}
\STATE $\mathcal{DIRECTOR} = Init\_Role(Prompt_{Dirctor}, \mathcal{M}) $
\STATE $\mathcal{EXECUTOR} = Init\_Role(Prompt_{Executor}, \mathcal{M}) $
\STATE \textcolor{black}{$Prompt_{Evaluator} = Prompt_{Evaluator}.join(K_{examples})$}
\STATE \textcolor{black}{ $\mathcal{EVALUATOR} = Init\_Role(Prompt_{Evaluator},  \mathcal{M})$}
\STATE \textcolor{gray}{\# Collaboration}
\STATE \textcolor{black}{Init: $epoch = 0, feedback = False$}
\STATE $ keypoint = \mathcal{DIRECTOR}(log) $
\STATE $ content = \mathcal{EXECUTOR}(log, keypoint) $
\REPEAT
\STATE \textcolor{black}{$ epoch = epoch + 1$}
\STATE \textcolor{black}{$ feedback = \mathcal{EVALUATOR}(log, keypoint, content) $}
\IF{\textcolor{black}{$feedback == False$}}
\STATE \textcolor{black}{$ content = \mathcal{EXECUTOR}(content, feedback) $}
\ENDIF
\UNTIL{\textcolor{black}{$epoch == Epoch$ or $feedback==True$}}
\STATE $k = content$
\RETURN $k$
\end{algorithmic}
\end{algorithm}

\begin{figure*}[]
\centering
   \includegraphics[width=\textwidth]{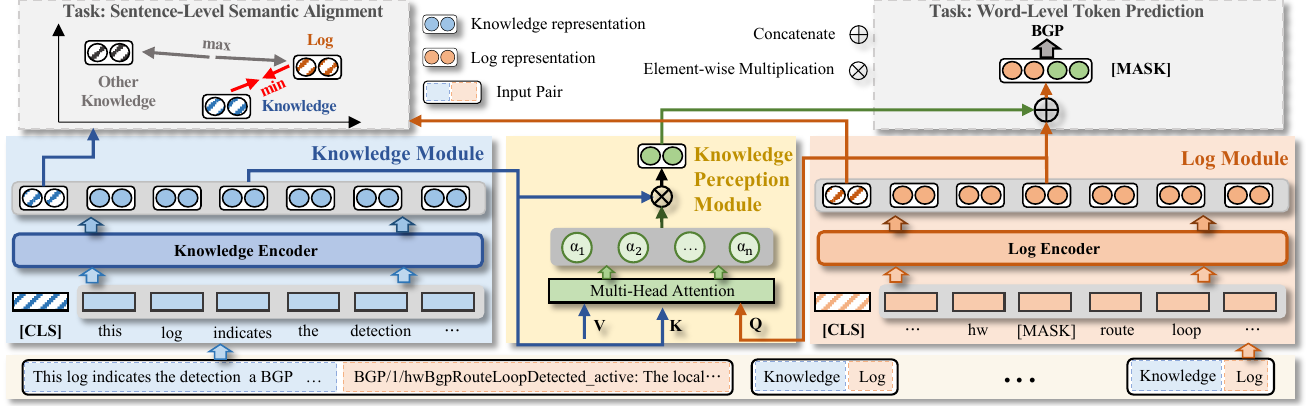}
   \vspace{-3mm}
   \caption{The overview of LUK pre-training framework. LUK includes two novel pre-training objectives on word level and sentence level. The word-level object is designed from the perspective of domain token understanding, while the sentence-level objective is designed from overall semantic alignment.}
   \label{fig:framework}
\end{figure*}

\subsection{Pre-Training with Knowledge Enhancement}
The overall pre-training framework is shown in Fig. \ref{fig:framework}, the input of the pre-training framework is a batch of pairs consisting of logs and the corresponding expert knowledge, then these pairs are fed into two encoders to obtain log representations and knowledge representations through two encoders respectively, \textcolor{black}{where two encoders are initialized with the same pre-trained model. } 
Finally, \textcolor{black}{
based on the structural characteristics of the log, we propose two pre-training tasks at word-level and sentence-level:} (1) token prediction and (2) semantic alignment, to incorporate expert knowledge for improving log understanding. 

%

\subsubsection{Word-Level Token Prediction}
Logs contain many domain terminologies, to sufficiently understand these domain terminologies, we propose a token prediction pre-training task (TP). Unlike existing word-level tasks of traditional PLMs only utilizing local context to predict tokens, our proposed word-level task requires models to aggregate context and knowledge for predicting tokens, leading to a knowledgeable pre-trained model.

Specifically, given a log $l$ and expert knowledge $k$ generated from the LLM, we get the corresponding input token sequence $ls= \{l_0,l_1,...,l_n\}, ks = \{k_0,k_1,...,l_m\}$ after tokenization. The token prediction task randomly masks a certain percentage (15\% in our experiments) of the log tokens $l_i$ with a special $[MASK]$ token, and then tries to recover them by perceiving external knowledge. To perceive knowledge, we design a knowledge perception module (KPM). The process of this module can be described in three steps:

Firstly, log encoder and knowledge encoder encode $ls$ and $ks$, respectively, and get the corresponding token representations $\boldsymbol{l} = \{\boldsymbol{l_0},\boldsymbol{l_1},...,\boldsymbol{l_n}\}$ and $\boldsymbol{k} = \{\boldsymbol{k_0},\boldsymbol{k_1},...,\boldsymbol{k_m}\}$.
Secondly, since not all tokens in knowledge contribute equally to the masked token prediction and to measure the importance of each token in knowledge for the token semantic, KPM calculates the semantic similarity between masked token $\boldsymbol{l_i}$ and knowledge $\boldsymbol{k}$, each token in knowledge is assigned a weight to represent its importance:
\begin{equation}
Q = W_Q\boldsymbol{l_i}, K = W_K\boldsymbol{k}, V = W_V\boldsymbol{k},
\end{equation}
\begin{equation}
\alpha = softmax(\frac{QK^T}{\sqrt{d_k}}),  \boldsymbol{k^{'}} = \alpha V,
\end{equation}
where $W_Q$, $W_K$ and $W_V$ are learnable parameter matrices, $d_k$ is the dimension of representation and $\alpha$ refers to the attention distribution. Thirdly, we concatenate the vector $\boldsymbol{l_i}$ of the masked token with $\boldsymbol{k^{'}}$ and use $[\boldsymbol{l_i};\boldsymbol{k^{'}}]$ to predict the original token:
\begin{equation}
\hat{y_i} = softmax(W_f[\boldsymbol{l_i};\boldsymbol{k^{'}}]),
\end{equation}
where $W_f$ is the weight parameter.
At last, the TP objective is to predict the original tokens which are masked out, formulated as follows:
\begin{equation}
    \mathcal{L}_{TP}(\theta) = -log \mathbf{p}(l_i | ls \backslash \{l_i\} ), \label{mlm}
\end{equation}
where $ls \backslash \{l_i\}$ denotes the log tokens sequence $ls$ with token $l_i$ being masked.

\subsubsection{Sentence-Level Semantic Alignment}
Since logs do not conform to the grammatical structure of human language and are always concise, it is difficult to sufficiently capture the semantics of logs. To enrich the semantic context of logs and improve the robustness of log representations, \textcolor{black}{inspired by KnowLog \cite{ma2024knowlog}, we construct a semantic alignment pre-training task (SA) in the sentence-level} that aligns the semantic representations of logs with the expert knowledge in the semantic space by contrastive learning.

Specifically, we first construct positive and negative examples from a batch $T = \{(l,k)\}$ of input pairs with size $N$. $(l^a,k^a)$ is the $a$-th pair in the batch. We obtain logs $l$ and knowledge $k$ from the same input pairs $\{(l^a,k^b)_{a=b}\}$ as positive examples and different input pairs $\{(l^a,k^b)_{a \neq b}\}$ as negative examples. Then we use $\boldsymbol{l^a}$, $\boldsymbol{k^a}$ to indicate the representations of the log and the knowledge, here the hidden state of the special symbol $[CLS]$ as the sentence representation. To pull closer positive samples and push away negative samples, the training objective of SA is defined as follows:
\begin{equation}
    f(\boldsymbol{l^a,k^b}) = exp(\frac{\boldsymbol{(l^a)}^{\top} \boldsymbol{k^b} }{\lVert \boldsymbol{l^a} \rVert \cdot \lVert \boldsymbol{k^b} \rVert} / \tau ), \label{ct}
\end{equation}

\begin{equation}
    \mathcal{L}_{SA}(\theta)=\frac{1}{N}\sum_{a=1}^N -log\frac{f(\boldsymbol{l^a,k^a})}{f(\boldsymbol{l^a,k^a}) + \sum\nolimits_{b=1}^{N-1} f(\boldsymbol{l^a,k^b})},
\end{equation}
where $\tau$ is the temperature hyper-parameter, which is set to 0.05 empirically in our experiments.

At last, we sum the TP loss and the SA loss, and obtain the overall training objective:
\begin{equation}
\min_{\theta} \mathcal{L}_{TP}(\theta) + \mathcal{L}_{SA}(\theta)
\end{equation}

\subsection{Fine-Tuning with Downstream Tasks}
After pre-training, we fine-tune the log encoder on different downstream tasks. We group all downstream tasks into two categories based on the input type: log-single tasks and log-pair tasks. We use the final hidden state of the first token (the $[CLS]$ token) as the sentence representation. For log-single tasks, we input the output of the language model $l_{CLS}$ into a multi-layer perception network function $f_H$ and obtain the prediction as $f_{H}(l_{CLS})$. For log-pair tasks, we follow sentence-bert \cite{reimers2019sentence} and use the language model to encode two separate sentences $u,v$, then input $u_{CLS},v_{CLS}$ and element-wise difference $|u_{CLS}-v_{CLS}|$ into a multi-layer perception network function $f_{H}$ and obtain the prediction as $f_{H}([u_{CLS};v_{CLS};|u_{CLS}-v_{CLS}|])$. 

\section{Experiments}

\subsection{Data preparation}
In this paper, we collect logs from software systems and network devices to construct expert knowledge and the log pre-trained model. Specifically, we collect software system logs from Loghub \cite{zhu2023loghub}, including operating system logs composed of Windows and BGL, as well as distributed system logs composed of HDFS and OpenStack.
Given that these datasets contain a significant proportion of duplicated logs, we employ widely used Drain \cite{he2017drain} to parse these logs and achieve log templates. Subsequently, we only sample one instance log as input that corresponds to each unique template.
In addition, we also collect network device logs from the public documentation of two vendors, Cisco\footnote{\url{https://www.cisco.com/c/en/us/support/all-products.html}} and Huawei\footnote{\url{https://support.huawei.com/enterprise/en/index.html}}, including three devices: Switches, Routers, and WLAN. The detailed statistics are shown in Table \ref{tab:statistics}.



\begin{table}[h]
\centering
\caption{Statistics of the dataset used for knowledge generation and pre-training.}
\label{tab:statistics}
\begin{tabular}{ll|c}
\toprule[1pt]
\multicolumn{1}{l|}{Datasets}                         & Category           & $\#$ of log templates \\ \midrule
\multicolumn{1}{l|}{\multirow{2}{*}{Software System}} & Distributed System & 2,292               \\
\multicolumn{1}{l|}{}                                 & Operating System   & 11,947              \\ \midrule
\multicolumn{1}{l|}{\multirow{2}{*}{Network Device}}  & Cisco              & 16,591              \\
\multicolumn{1}{l|}{}                                 & Huawei             & 12,399              \\ \midrule
\multicolumn{2}{c|}{Total}                                                 & 43,229              \\ \bottomrule[1pt]
\end{tabular}%
\end{table}

\begin{figure*}[]
\centering
   \includegraphics[width=\textwidth]{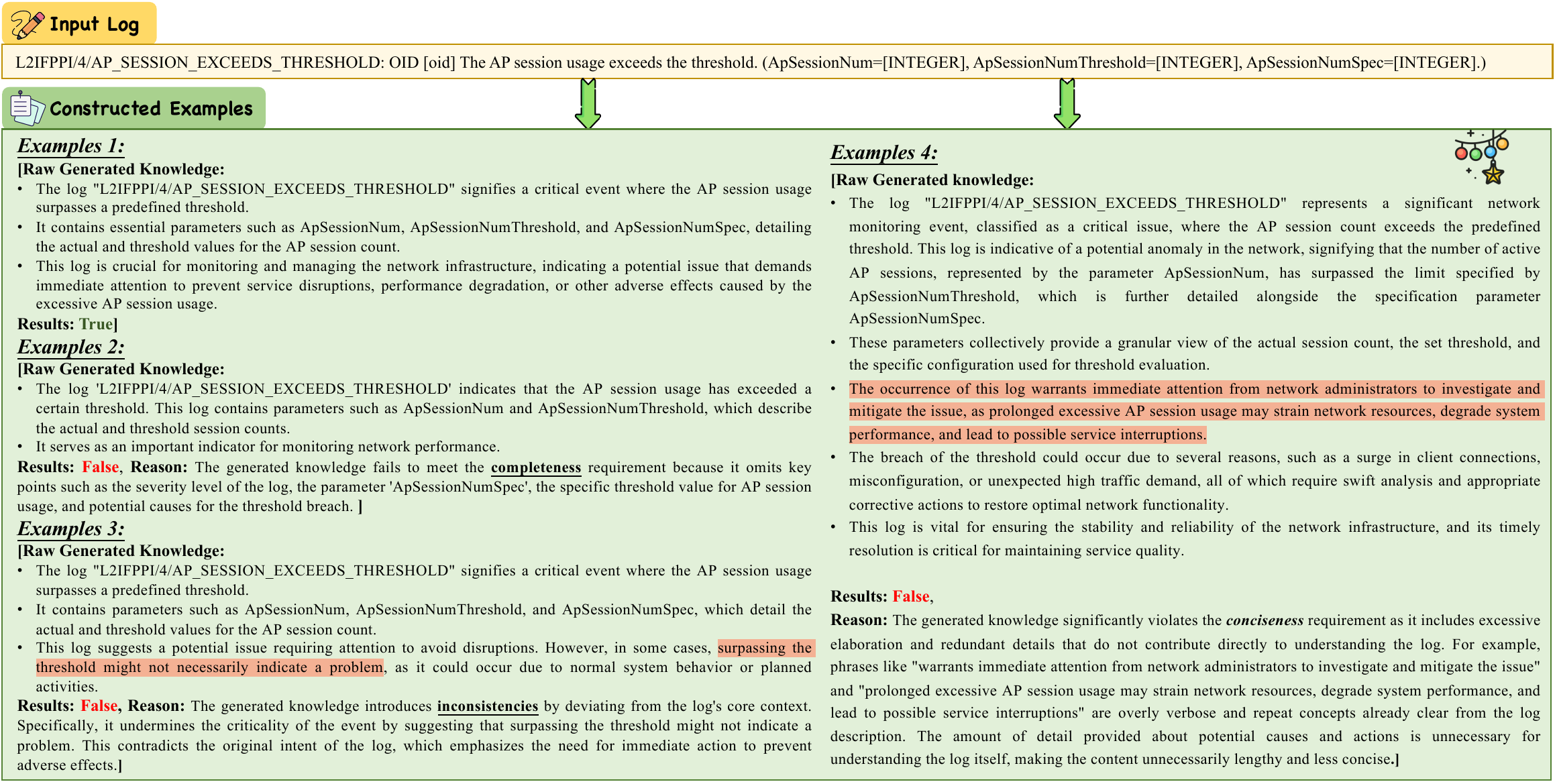}
   \vspace{-3mm}
   \caption{\textcolor{black}{The contrastive samples constructed in our experiments, where Example 1 represents the positive example, and the remaining three examples are negative examples obtained from the modification of GPT-4o. Colour-marked text indicates that the evaluation requirements are unsatisfied. }}
   \vspace{-3mm}
   \label{fig:evaluation}
\end{figure*}

\subsection{Parameters Setting}
\begin{itemize}
\item  \textcolor{black}{As for expert knowledge acquisition, we explore different large language models for experiments, including two proprietary LLMs: \textit{gpt-3.5-turbo} (ChatGPT) and \textit{gpt-4o} (GPT-4o), and one open-source model: \textit{Llama-3.3-70B-Instruct } (LLama3-70B) \cite{dubey2024llama}.} To increase the stability of LLM’s output, we set the temperature of LLMs to 0.
\item \textcolor{black}{
For contrastive examples provided to the \textit{Evaluator}, we show in Fig. \ref{fig:evaluation}.}
\item For pre-training, we use \textit{bert-base-uncased} with 110M parameters in our experiments. During pre-training, we set the batch size as 32, epochs as 40 and the maximum length of the input text as 512. Moreover, the optimizer we adopt is Adam with a learning rate of 5e-5, a weight decay of 0.01, learning rate warmup for 2,000 steps and linear decay of the learning rate after.
\item For fine-tuning, we adopt the cross-entropy loss as the loss function in downstream tasks and we set epoch to 20 and 10 on log-single and log-pair tasks, respectively.
\item \textcolor{black}{We spend 3 hours pre-training the LUK on 2 NVIDIA A100 GPUs with PyTorch 2.4.0.}
\end{itemize}

\begin{table}[]
\centering
\caption{Statistics of downstream tasks datasets on software system  (Training/Validation/Testing Size).}
\label{tab:software_statistics}
\begin{tabular}{llcll}
\toprule[1pt]
\multicolumn{1}{l|}{Tasks}                                                                        & Dataset      & \multicolumn{3}{c}{\#Size}        \\ \midrule
\multicolumn{1}{l|}{\multirow{3}{*}{\begin{tabular}[c]{@{}l@{}}Anomaly\\ Detection\end{tabular}}} & BGL          & \multicolumn{3}{c}{30K / 10K / 10K} \\
\multicolumn{1}{l|}{}                                                                             & ThunderBird* & \multicolumn{3}{c}{30K / 10K / 10K} \\
\multicolumn{1}{l|}{}                                                                             & Spirit*      & \multicolumn{3}{c}{30K / 10K / 10K} \\ \midrule
\multicolumn{1}{l|}{\begin{tabular}[c]{@{}l@{}}Failure \\ Identification\end{tabular}}            & Openstack    & \multicolumn{3}{c}{236 / 80 / 80} \\ \bottomrule[1pt]
\multicolumn{5}{l}{* indicates logs not involved in the pre-training.}                                                                            
\end{tabular}%
\end{table}

\begin{table}[]
\caption{Statistics of downstream tasks datasets on network device (Training/Validation/Testing Size).}
\tabcolsep = 0.07cm
\label{tab:network_statistics}
\resizebox{\columnwidth}{!}{%
\begin{tabular}{lllll}
\toprule[1pt]
\multicolumn{1}{l|}{Tasks}                                                                                            &        & Switches             & Routers            & Security*         \\ \midrule
\multicolumn{1}{l|}{\multirow{3}{*}{\begin{tabular}[c]{@{}l@{}}Module\\ Classification\end{tabular}}}                 & Cisco  & 13,495/4,498/4,498   & 7,265/2,422/2,421  & -                 \\
\multicolumn{1}{l|}{}                                                                                                 & Huawei & 3,439/1,146/1,146    & 2,539/846/845      & -                 \\
\multicolumn{1}{l|}{}                                                                                                 & H3C*   & 1,241/413/413        & 1,336/445/444      & -     \\ \midrule
\multicolumn{1}{l|}{\begin{tabular}[c]{@{}l@{}}Fault Phenomenon\\ Identification\end{tabular}}                        & Huawei & 362/120/120          & -                  & -                 \\ \midrule
\multicolumn{1}{l|}{\multirow{3}{*}{\begin{tabular}[c]{@{}l@{}}Log and Description\\ Semantic Matching\end{tabular}}} & Cisco  & 49,954/16,651/16,651 & 26,975/8,992/8,991 & 1,894/631/631     \\
\multicolumn{1}{l|}{}                                                                                                 & Huawei & 7,702/2,567/2,567    & 5,977/1,992/1,991  & 4,485/1,495/1,494 \\
\multicolumn{1}{l|}{}                                                                                                 & H3C*   & 2,606/868/868        & 2,837/946/945      & 2,223/741/740     \\ \midrule
\multicolumn{1}{l|}{\begin{tabular}[c]{@{}l@{}}Log and Possible\\ Cause Ranking\end{tabular}}                         & Huawei & 3,851/1,283/1,283    & 3,097/1,032/1,032  & 2,361/787/787     \\ \bottomrule[1pt]
\multicolumn{5}{l}{* indicates logs not involved in the pre-training.}                                                                                                                        
\end{tabular}%
}
\end{table}

\subsection{Downstream tasks}
To explore the performance of LUK on different log analysis domains, we conduct experiments on different downstream tasks, including software system and network device logs. 

Firstly, following existing studies in software system logs \cite{zhu2023loghub,he2021survey}, we evaluate LUK on two widely researched log analysis tasks to validate the representativeness of our approach: anomaly detection and failure identification. 

Secondly, we conduct experiments on network device logs to investigate the model's generalization ability. This domain poses additional challenges due to the presence of various vendors and highly specialized logs. Due to the lack of public tasks on network device logs, we construct four different downstream tasks referring to \cite{ma2024knowlog}. 
These tasks are commonly encountered by engineers and require substantial expert knowledge for effective solutions.

\textcolor{black}{
The evaluation datasets in our experiments include two types: including in the pre-training corpus and not including in the pre-training corpus. For most datasets, we collect the evaluation datasets from the same source as the pre-training data.
Moreover, 
apart from logs involved in pre-training, to verify the generalization capability of LUK, we collect more logs that are not included in the pre-training for evaluation.}
In Table \ref{tab:software_statistics} - \ref{tab:network_statistics} we provide statistics for different tasks of their datasets. Next, we give an introduction to each task and its evaluation metrics.
\subsubsection{Downstream Tasks of Software System Logs}

\begin{itemize}
    \item \textbf{Anomaly Detection (AD).} Anomaly detection is a widely researched log analysis task to predict whether anomalies exist within a short period of log messages, where the input is a log sequence and the output is True or False.  Following Biglog \cite{tao2023biglog}, we concatenate each log in the sequence and then input them to the encoder to obtain the representation of the sequence.
    
    \textbf{Dataset and Metric.} Following previous anomaly detection studies \cite{le2021log,le2022log}, we collect datasets from Loghub \cite{zhu2023loghub}. To measure the effectiveness of different models in anomaly detection, we report \textit{Precision}, \textit{Recall}, and \textit{F1} on the True (anomaly) class as evaluation metrics.
    
    \item \textbf{Failure Identification (FI).} Failure identification aims to further discern the type of failure present in the anomaly log. Given the log messages, the model is required to determine what error emerges. 

    \textbf{Dataset and Metric.} This dataset comes from \cite{cotroneo2019bad}, which is an OpenStack dataset including 396 failure tests and 16 kinds of API errors, such as ``network delete error'', ``openstack network create error''. Usually, engineers are interested in whether top-K recommended results contain the correct error, hence, we report the \textit{Recall@K} rate as the evaluation metric.
    
\end{itemize}

\begin{table*}[]
\centering
\caption{Examples of Module Classification (MC) and Fault Phenomenon Identification (FPI).}
\label{tab:tasks}
\resizebox{0.85\textwidth}{!}{%
\begin{tabular}{l|cl}
\toprule[1pt]
Tasks                & \multicolumn{2}{c}{Example}                                                                                                                                                                                                                                                                                                                                          \\ \midrule
\multirow{4}{*}{MC}  & \textbf{Input}  & \begin{tabular}[c]{@{}l@{}}{[}MASK{]}/6/NOTIFY\_RECV: The router received a NOTIFICATION from the peer. (Peer={[}peer-address{]}, \\ SourceInterface={[}SourceInterface{]}, ErrorCode={[}error-code{]}, SubErrorCode={[}sub-error-code{]}, NotifyInfo={[}notify-info{]},\\ VpnInstance={[}VpnInstance{]}, ErrorData={[}error-data{]})\end{tabular} \\
                     & \textbf{Output} & BGP                                                                                                                                                                                                                                                                                                                                                \\ \cmidrule{2-3} 
                     & \textbf{Input}  & {[}MASK{]}-3-DUPLICATE\_IFINDEX:\%s has \%d duplicate ifIndices.                                                                                                                                                                                                                                                                                   \\
                     & \textbf{Output} & SNMP  \\ \midrule
\multirow{4}{*}{FPI} & \textbf{Input}  & \begin{tabular}[c]{@{}l@{}}OSPF/4/CONFLICT\_ROUTERID\_INTF: OSPF router ID conflict is detected on the interface.(ProcessId=1, RouterId=\\ 10.26.09.101, AreaId=0.0.0.0, InterfaceName=10GE1/0/11, IpAddr=10.26.10.1, PacketSrcIp=10.26.10.2)\end{tabular}                                                                                         \\
                     & \textbf{Output} & Router\_id\_conflict                                                                                                                                                                                                                                                                                                                               \\ \cmidrule{2-3} 
                     & \textbf{Input}  & \begin{tabular}[c]{@{}l@{}}IFNET/2/linkDown\_activ: The interface status changes. (ifName=10GE1/0/11, AdminStatus=DOWN,  OperStatus=\\ DOWN, Reason=The interface is shut down, mainIfname=10GE1/0/11)\end{tabular}                                                                                                                                \\
                     & \textbf{Output} & Trunk\_link\_down \& Physical\_link\_down                                                                                                                                                                                                                                                                                                          \\ \bottomrule[1pt]
\end{tabular}%
}
\end{table*}

\begin{table*}[]
\centering
\caption{Examples of Log and Description Semantic Matching (LDSM) and Log and Possible Cause Ranking (LPCR).}
\label{tab:tasks2}
\resizebox{0.85\textwidth}{!}{%
\begin{tabular}{c|cl}
\toprule[1pt]
\multicolumn{1}{l|}{Tasks}                 & \multicolumn{2}{c}{Example}                                                                                                                                                                                                                                                                                                     \\ \midrule
\multirow{4}{*}{LDSM}                      & \textbf{Input}  & \begin{tabular}[c]{@{}l@{}}{[} ARP/4/ARP\_VLAN\_SPEED\_LMT: The VLAN's ARP packet speed exceeded the configured speed limit value. \\ (SuppressValue={[}SpeedLmtValue{]}, Vlan={[}VlanId{]}) , \\ The transmit rate of ARP packets in a VLAN exceeded the configured rate limit in the VLAN. {]}\end{tabular} \\
                                           & \textbf{Output} & True                                                                                                                                                                                                                                                                                                          \\ \cmidrule{2-3} 
                                           & \textbf{Input}  & \begin{tabular}[c]{@{}l@{}}{[}(ARP/4/ARP\_VLAN\_SPEED\_LMT: The VLAN's ARP packet speed exceeded the configured speed limit value. \\ (SuppressValue={[}SpeedLmtValue{]}, Vlan={[}VlanId{]}) , \\ A received ARP packet was not an ARP reply packet in response to the ARP request packet sent by the device. 
 {]}\end{tabular}       \\
                                           & \textbf{Output} & Fasle                                                                                                                                                                                                                                                                                                         \\ \midrule
\multirow{4}{*}{LPCR}                      & \textbf{Input}  & \begin{tabular}[c]{@{}l@{}}BGP/3/FSM\_UNEXPECT: FSM received an unexpected event. (FSM={[}fsm-name{]}, PreState={[}prev-state{]}, \\ CurrState={[}curr-state{]}, InputEvent={[}input{]})\end{tabular}                                                                                                         \\
                                           & \textbf{Output} & It is caused by an internal error of the system.                                                                                                                                                                                                                                                              \\ \cmidrule{2-3} 
                                           & \textbf{Input}  & \begin{tabular}[c]{@{}l@{}}BGP/2/hwBgpPeerSessionExceed\_clear: The number of BGP peer sessions decreased below the maximum number. \\ (MaximumNumber={[}MaximumNumber{]}, CurrentNumber={[}CurrentNumber{]})\end{tabular}                                                                                    \\
                                           & \textbf{Output} & The number of BGP peer sessions fell below the upper limit.                                                       \\ \bottomrule[1pt]
\end{tabular}%
}
\end{table*}

\subsubsection{Downstream Tasks of Network Device Logs}

To intuitively understand these tasks, we give examples of each task in Table \ref{tab:tasks} - \ref{tab:tasks2}.
\begin{itemize}
    \item \textbf{Module Classification (MC).} MC is a log-single task aiming at identifying which module the log originates from, which input is a log with the masked module name and the output is the corresponding module name. 

    \textbf{Dataset and Metric.} We collect the log from Table \ref{tab:network_statistics} and replace the module name with \textit{[MASK]} as input, the module name in the log as ground truth. Obviously, this is a multi-classification task and the model needs to understand the contextual information of the logs to accurately identify their source.
    As an unbalanced multi-class classification task and considering the importance of different classes, we report \textit{Accuracy} and \textit{Weighted F1} as evaluation metrics.
    
    \item \textbf{Fault Phenomenon Identification (FPI).} FPI is also a log-single task to identify the fault category to which the log belongs. Different from the previous failure identification task, FPI is a multi-label classification task due to a log may appear in more than one fault category.
    
    \textbf{Dataset and Metric.} We collect 602 Huawei switches logs covering 43 fault categories from real-world as the dataset, these logs are annotated by experts. Unlike the multi-class classification task, we report \textit{Average Accuracy} of all samples \cite{sorower2010literature} as the evaluation metric, where the accuracy for each sample is the number of correctly predicted labels.

    \item \textbf{Log and Description Semantic Matching (LDSM).} LDSM is a log-pair task aimed at determining whether the semantics of a given log align with the corresponding natural language description, where the input is a log and description pair, and the output is True or False.

    \textbf{Dataset and Metric.} We collect descriptions of logs from the documentation, then we build (log, description) pair as ground truth and randomly select one other description for each log as a negative sample. This task requires the model to accurately understand the semantics of the logs and descriptions. As a binary classification task, both positive and negative cases require attention, we report the \textit{Accuracy} and \textit{Weighed F1} as evaluation metrics.
    
    \item \textbf{Log and Possible Cause Ranking (LPCR).} LPCR is a log-pair ranking task to find the most probable answer from a list of possible causes for a given log. \textcolor{black}{
    Specifically, the input is a log and an answer candidate set, then the model needs to find the most probable causes from the set and rank the candidate set based on the priority as the output. }

    \textbf{Dataset and Metric.} We collect logs and the corresponding possible causes from the Huawei public documentation and build (log, possible cause) pairs as ground truth. Then we randomly select 15 possible causes of other logs the ground truth as a candidate set. This task necessitates that the model understands the background information of the log to accurately identify its potential causes. As a typical ranking task, following \cite{niu2022spt}, we report \textit{Precision@K} and \textit{Mean Reciprocal Rank} (MRR) as evaluation metrics, where MRR is a statistic measure for evaluating search algorithms. 
\end{itemize}

\subsection{Baselines}

\textcolor{black}{
We categorize baselines for log understanding into four groups according to technology type: traditional deep-learning methods, pre-trained language models, knowledge distillation methods, and large language models. In addition, to ensure a fair comparison, apart from LLMs, we re-produce all baselines from their repositories, and the parameters of baseline models are according to their original settings. } 
\subsubsection{\textcolor{black}{Traditional deep-learning methods}} 
\textcolor{black}{Traditional deep-learning methods convert each log message into a vector with the word embedding model and then input the vector to a deep neural network (CNN or BiLSTM) to analyze logs. Following previous works \cite{zhang2019robust, lu2018detecting}, We choose the most widely used \textbf{BiLSTM} and \textbf{CNN} for log analysis as deep neural network building models.}

\subsubsection{\textcolor{black}{Pre-trained language models}}
\textcolor{black}{Pre-trained language model-based log analysis methods follow the \textit{pre-train and fine-tune} paradigm, where one pre-trained language model serves as the backbone, and then is fine-tuned on different specific downstream tasks. We select one general pre-trained model (\textbf{BERT \cite{le2021log,lee2023lanobert}}) and two specific pre-trained models for logs (\textbf{Biglog \cite{tao2023biglog}} and \textbf{KnowLog \cite{ma2024knowlog}}) as baselines.
\textbf{Biglog} excels in capturing essential features due to its extensive training on the log corpus. In addition, \textbf{KnowLog} is the state-of-the-art log pre-trained model in log analysis not only due to pre-training on log corpus but also enhancing the pre-training with documentation knowledge. Referring to KnowLog, we collect the descriptions of logs from the Huawei and Cisco public
documentation as background knowledge to enhance the log
pre-training in the same way.
For fairness, we reproduce Biglog and KnowLog with the same pre-training setting on our log corpus. 
}

\subsubsection{\textcolor{black}{Knowledge distillation methods}}
\textcolor{black}{
In the era of LLMs, data augmentation (DA) is the prevalent paradigm to achieve knowledge distillation of LLMs, where it prompts the LLM to generate more data tailored to train the student model in a targeted skill or domain \cite{taori2023stanford,xu2024survey}. This reasoning process-oriented distillation method is more effective compared to obtaining labels directly from the teacher model \cite{zhu2024distilling}.
Considering the lack of research on knowledge distillation in the field of log analysis, we select Chain-of-Thought (COT)-based knowledge distillation as the baseline, which has proven to be an effective knowledge distillation method for the reasoning process in the field of NLP \cite{hsieh2023distilling,shridhar2023distilling}.
}

\textcolor{black}{
Specifically, we select \textbf{T5 \cite{raffel2020exploring}} (\textit{t5-base with 223M parameters})  as the student model, and GPT-4o as the teacher model.  Given the specific log analysis task, we provide the inputs and labels from the training set to the teacher model, then utilize COT prompting to extract rationales from the teacher model. Finally, the student model is trained with the inputs and constructed rationales. }

\subsubsection{\textcolor{black}{Large language models}}
\textcolor{black}{LLM-based log analysis methods follow the \textit{in-context learning} paradigm, where LLMs make predictions based on contexts augmented with a few examples without training.
Following previous LLM-based log analysis works \cite{xu2024divlog,xu2024unilog,liu2023logprompt,jiang2023llmparser}, we select top-5 similar examples from the training set for each input log and use \textit{bge-large-en-v1.5} as the embedding model to select examples, due to its excellent performance among all open-source embedding models. Considering the recency bias of LLMs \cite{zhao2021calibrate}, we arrange these examples in ascending order based on the cosine similarity. We select two proprietary LLMs (\textbf{ChatGPT} and \textbf{GPT-4o}) and one powerful open-source LLM (\textbf{LLama3-70B \cite{dubey2024llama}}) as baselines. We utilize HTTP requests to invoke the OpenAI APIs and interact with proprietary LLMs. For open-source LLM, we deploy \textit{Llama-3.3-70B-Instruct} in our environments to analyze logs.}

In addition, since \textbf{COT} prompting \cite{wei2022chain} is verified to improve the performance of LLMs significantly. To verify the effectiveness of multi-expert collaboration, we acquire knowledge with COT and then train the model with the same pre-training tasks for comparison. 
The specific prompt is: \textit{You are an engineer in the maintenance and operation domain, Please help me understand this log, including parameters, description, possible causes and resolution procedures. Let's think step by step.}

\begin{table}[]
\color{black}

\centering
\caption{Results on AD and FI (software system logs).}
\label{tab: software_results}
\resizebox{\columnwidth}{!}{%
\begin{tabular}{lc|c}
\toprule[1.5pt]
\multirow{2}{*}{Methods} & AD (Precision / Recall / F1) & FI (Recall@1 / 2 / 3)                     \\
                         & BGL                          & OpenStack                                 \\ \midrule
CNN                      & \textbf{99.95} / 96.06 / 97.97        & 83.75 / 86.25 / 90.00                     \\
BiLSTM                   & \textbf{99.95} / 97.25 / 98.58        & 87.50 / 92.50 / 96.25                     \\
BERT                     & 99.70 / 98.48 / 99.09        & 87.50 / 92.50 / \textbf{98.75}                     \\
Biglog                   & 99.87 / 99.63 / 99.75        & 90.00 / 95.00 / \textbf{98.75}                     \\
KnowLog                  & 99.58 / 98.69 / 99.13        & 82.50 / 92.50 / 96.25                     \\ \midrule
T5-KD                    & 99.87 / 99.56 / 99.91        & 86.25 / 92.50 / 98.75                     \\ \midrule
LLama3-70B               & 46.87 / 95.23 / 71.05        & 85.00 / \textbf{96.25} / 96.25                     \\
ChatGPT                  & 42.73 / 95.23 / 68.98        & 83.75 / 95.00 / 96.25                     \\
GPT-4o                   & 46.85 / 95. 23 / 70.54       & 87.50 / \textbf{96.25} / 96.25                     \\ \midrule
LUK (COT-ChatGPT)        & 99.95 / 99.87 / 99.91        & 91.25 / 95.00 / \textbf{98.75}                     \\
LUK (COT-GPT4o)          & 99.67 / \textbf{100.0} / 99.83        & 91.25 / 95.00 / \textbf{98.75}                     \\ \midrule
LUK (MEC-LLama3)         & 99.87 / \textbf{100.0} / \textbf{99.93}        & \multicolumn{1}{l}{\textbf{92.50} / 95.00 / \textbf{98.75}} \\
LUK (MEC-ChatGPT)        & 99.95 / 99.91 / 99.93        & \multicolumn{1}{l}{\textbf{92.50} / 95.00 / \textbf{98.75}} \\
LUK (MEC-GPT4o)          & 99.87 / \textbf{100.0} / \textbf{99.93}       & \multicolumn{1}{l}{\textbf{92.50} / \textbf{96.25} / 96.25} \\ \bottomrule[1.5pt]
\end{tabular}%
}
\end{table}


\begin{table*}[]
\color{black}
\centering
\caption{Results on Module Classification (MC) and Log and Description Semantic Matching (LDSM) tasks. (network device logs).}
\label{tab: network_result}
\resizebox{\textwidth}{!}{%
\begin{tabular}{lcccc|cccc}
\toprule[1.5pt]
\multirow{3}{*}{Methods} & \multicolumn{4}{c|}{MC (Accuracy / Weighted F1)}                                   & \multicolumn{4}{c}{LDSM (Accuracy / Weighted F1)}                                  \\ \cmidrule{2-9} 
                         & \multicolumn{2}{c|}{Cisco}                         & \multicolumn{2}{c|}{Huawei}   & \multicolumn{2}{c|}{Cisco}                         & \multicolumn{2}{c}{Huawei}    \\
                         & Switches      & \multicolumn{1}{c|}{Routers}       & Switches      & Routers       & Switches      & \multicolumn{1}{c|}{Routers}       & Switches      & Routers       \\ \midrule
CNN                      & 56.89 / 56.85 & \multicolumn{1}{c|}{57.46 / 54.92} & 74.52 / 73.95 & 72.78 / 72.23 & 84.04 / 84.04 & \multicolumn{1}{c|}{80.99 / 80.99} & 86.05 / 86.05 & 82.37 / 82.30 \\
BiLSTM                   & 55.74 / 55.63 & \multicolumn{1}{c|}{57.17 / 56.76} & 76.52 / 75.49 & 73.96 / 73.30 & 89.45 / 89.44 & \multicolumn{1}{c|}{85.42 / 85.41} & 87.85 / 87.85 & 84.43 / 84.40 \\
BERT                     & 62.67 / 61.38 & \multicolumn{1}{c|}{62.72 / 62.60} & 82.37 / 81.20 & 81.18 / 79.20 & 93.06 / 93.06 & \multicolumn{1}{c|}{90.01 / 90.00} & 93.18 / 93.18 & 90.06 / 90.05 \\
Biglog                   & 62.69 / 62.76 & \multicolumn{1}{c|}{63.15 / 61.17} & 83.24 / 83.25 & 82.36 / 81.19 & 93.32 / 93.32 & \multicolumn{1}{c|}{91.46 / 91.46} & 94.19 / 94.19 & 93.62 / 93.61 \\
KnowLog                & 63.33 / 63.35 & \multicolumn{1}{c|}{63.65 / 63.12} & 84.11 / 83.65 & 83.55 / 82.41 & 95.05 / 95.05 & \multicolumn{1}{c|}{92.78 / 92.78} & 97.15 / 97.15 & 96.43 / 96.43 \\ \midrule
T5-KD                    & 62.96 / 62.36 & \multicolumn{1}{c|}{63.86 / 61.01} & 83.60 / 83.23 & 83.55 / 82.82 & 95.29 / 95.28 & \multicolumn{1}{c|}{93.20 / 93.20} & 95.25 / 95.25 & 93.97 / 93.97 \\ \midrule
LLama3-70B               & 61.61 / 60.95 & \multicolumn{1}{c|}{62.99 / 60.55} & 82.98 / 82.83 & 81.66 / 80.34 & 84.48 / 84.31 & \multicolumn{1}{c|}{88.32 / 88.24} & 95.49 / 95.49 & 95.57 / 95.57 \\
ChatGPT                  & 57.11 / 54.41 & \multicolumn{1}{c|}{57.99 / 54.56} & 79.14 / 77.19 & 78.22 / 75.51 & 82.14 / 81.96 & \multicolumn{1}{c|}{84.64 / 84.47} & 85.75 / 85.75 & 90.31 / 90.22 \\
GPT-4o                   & 62.12 / 61.62 & \multicolumn{1}{c|}{63.86 / 61.01} & 83.25 / 82.80 & 81.30 / 79.87 & 89.04 / 88.99 & \multicolumn{1}{c|}{93.12 / 93.11} & 97.23 / 97.23 & \textbf{97.79} / \textbf{97.79} \\ \midrule
LUK (COT-ChatGPT)        & 63.49 / 63.72 & \multicolumn{1}{c|}{64.51 / 63.33} & 83.42 / 82.70 & 83.19 / 82.34 & 94.93 / 94.92 & \multicolumn{1}{c|}{93.11 / 93.11} & 96.33 / 96.33 & 95.73 / 95.73 \\
LUK (COT-GPT4o)          & 63.47 / 63.51 & \multicolumn{1}{c|}{63.61 / 63.02} & 83.60 / 83.20 & 82.84 / 81.92 & 94.92 / 94.91 & \multicolumn{1}{c|}{93.18 / 93.18} & 96.46 / 96.46 & 95.93 / 95.93 \\ \midrule
LUK (MEC-LLama3)         & 65.67 / 65.27 & \multicolumn{1}{c|}{66.67 / 65.12} & 85.25 / 84.95 & 85.21 / 84.01 & 96.45 / 96.45 & \multicolumn{1}{c|}{94.13 / 94.13} & 97.66 / 97.66 & 97.39 / 97.39 \\
LUK (MEC-ChatGPT)        & 65.54 / 65.12 & \multicolumn{1}{c|}{66.29 / 64.27} & 84.64 / 84.47 & 85.09 / 83.82 & 96.32 / 96.32 & \multicolumn{1}{c|}{93.80 / 93.80} & 96.92 / 96.92 & 96.74 / 96.74 \\
LUK (MEC-GPT4o)          & \textbf{66.16} / \textbf{66.02} & \multicolumn{1}{c|}{\textbf{67.20} / \textbf{65.41}} & \textbf{85.51} / \textbf{85.26} & \textbf{85.33} / \textbf{84.73} & \textbf{96.63} / \textbf{96.63} & \multicolumn{1}{c|}{\textbf{94.32} / \textbf{94.32}} & \textbf{97.66} / \textbf{97.66} & 97.34 / 97.34 \\ \bottomrule[1.5pt]
\end{tabular}%
}
\end{table*}

\begin{table}[]
\color{black}

\centering
\caption{Results on LPCR and FPI (network device logs). Note: The results of proprietary LLMs on the FPI task cannot be compared due to data privacy issues.}
\label{tab: network_result2}
\resizebox{\columnwidth}{!}{%
\begin{tabular}{lcc|c}
\toprule[1.5pt]
\multirow{3}{*}{Methods} & \multicolumn{2}{c|}{LPCR (Precision@1 / 3 / MRR)} & FPI (Accuracy) \\ \cmidrule{2-4} 
                         & \multicolumn{2}{c|}{Huawei}                       & Huawei         \\
                         & Switches                & Routers                 & Switches       \\ \midrule
CNN                      & 54.30 / 77.26 / 67.99   & 53.45 / 75.77 / 67.35   & 69.00          \\
BiLSTM                   & 59.27 / 78.04 / 71.22   & 51.45 / 69.56 / 63.76   & 74.83          \\
BERT                     & 76.18 / 91.54 / 84.70   & 72.57 / 91.59 / 82.61   & 88.75          \\
Biglog                   & 83.33 / 94.02 / 89.38   & 82.98 / 95.59 / 89.49   & 89.58          \\
KnowLog                  & 87.50 / 93.01 / 91.84   & 89.68 / 96.59 / 93.99   & 94.17          \\ \midrule
T5-KD                    & 86.11 / 93.56 / 91.41   & 84.48 / 92.59 / 90.33   & -              \\ \midrule
LLama3-70B               & \textbf{95.61} / \textbf{98.04} / \textbf{97.27}   & \textbf{97.48} / \textbf{99.49} / \textbf{98.52}   & 88.83          \\
ChatGPT                  & 90.95 / 95.11 / 93.88   & 97.20 / 98.34 / 98.19   & -              \\
GPT-4o                   & 95.65 / 98.29 / 97.27   & 97.39 / 99.09 / 98.47   & -              \\ \midrule
LUK (COT-ChatGPT)        & 85.18 / 95.03 / 90.55   & 84.58 / 95.99 / 90.64   & 90.42          \\
LUK (COT-GPT4o)          & 86.34 / 93.79 / 91.37   & 87.88 / 95.19 / 92.75   & 92.50          \\ \midrule
LUK (MEC-LLama3)         & 90.45 / 95.81 / 94.06   & 90.49 / 96.99 / 94.52   & 95.83          \\
LUK (MEC-ChatGPT)        & 89.37 / 95.19 / 93.43   & 88.48 / 94.69 / 90.33   & 93.33          \\
LUK (MEC-GPT4o)          & 90.92 / 96.04 / 94.46   & 90.69 / 96.09 / 94.39   & \textbf{96.67}          \\ \bottomrule[1.5pt]
\end{tabular}%
}
\end{table}

\subsection{Evaluation}
We evaluate LUK by answering the following research questions (RQs):

\begin{itemize}
    \item \textbf{RQ1: How effective is LUK compared with the current mainstream methods on downstream tasks?}
\end{itemize}
In this RQ, we conduct a comprehensive evaluation of LUK in comparison to other state-of-the-art baselines on software systems and network device logs. 

The experiment results are shown in Table \ref{tab: software_results} and \ref{tab: network_result2}, 
\textcolor{black}{it is clear that apart from the LLMs-based baselines, LUK outperforms all other baselines. 
In addition, LUK with smaller parameters outperforms the LLMs on all tasks except for the MC and LDSM tasks.
In particular, compared to the best results of baselines on the Failure Identification task of software system logs, LUK improves on Recall@1 by 2.5\%. And on the Module Classification task of network device logs, LUK's accuracy improved by an average of 2.34\%. 
This evidence underscores the superior performance of LUK in the domain of log analysis.}
Further, traditional DL models exhibit relatively poor performance, primarily attributed to their limited capacity to capture the semantics of logs. 
When compared to BERT and Biglog, \textcolor{black}{it becomes evident that pre-training on extensive log corpus enhances the semantic understanding abilities of PLMs. However, a more in-depth analysis uncovers a notable challenge: the presence of domain-specific terminologies and the concise nature of logs act as impediments, and the lack of introduction of external knowledge limited these models from achieving further breakthroughs.}

\textcolor{black}{To verify the effectiveness of
the expert knowledge obtained from the LLM, we compare
it with the KnowLog, whose knowledge is obtained from the documentation. Our experimental findings demonstrate that MEC-based LUK surpasses KnowLog in performance, more specifically, LUK (MEC-GPT4o) improves the average Accuracy on the Cisco and Huawei datasets regarding the MC and LDSM tasks by 2.38\% and 1.15\%, respectively.  
It is noteworthy that Huawei's documentation provides more detailed log descriptions compared to the relatively concise descriptions found in Cisco's documentation. This gap underscores the LLM's capability to generate expert knowledge of equal or superior quality to that found in high-quality documentation.
Given the scarcity of documented descriptions for most logs in practice, obtaining expert knowledge through KnowLog becomes impractical. For instance, the challenge of sourcing relevant documents for logs from sources like Loghub \cite{zhu2023loghub} significantly hampers the effectiveness of KnowLog. Our approach, which operates independently of documentation, consistently yields superior results and serves to diminish the reliance on human experts for intricate knowledge construction processes, offering a more feasible and practical alternative to traditional documentation-based methodologies. 
}

\textcolor{black}{In comparison to the knowledge distillation-based method (T5-KD), the chain-of-thought knowledge distillation approach exhibited suboptimal performance in specific log analysis tasks. 
This gap can be attributed to the unique nature of log analysis, which typically necessitates a deep understanding of domain-specific knowledge. The chain-of-thought methodology, with its sequential reasoning framework, may inadvertently introduce noise or overlook the analysis of relevant expert knowledge crucial for these tasks, leading to diminished performance in tasks where precise understanding and retention of domain-specific information are crucial. 
In comparison, LUK prioritizes the construction of high-quality expert knowledge explicitly before fine-tuning it for specific tasks. LUK is better positioned to address the intricate challenge inherent in log analysis tasks by establishing a strong foundation of domain-specific knowledge upfront. }

\textcolor{black}{
In comparison to LLMs, our results show that apart from LDSM and LPCR tasks, three LLMs perform weaker than the fine-tuned smaller PLMs. This disparity can be ascribed to the presence of a significant domain gap, which hampers the ability of LLMs to effectively address specialized tasks directly. Consequently, this limitation restricts their potential to fully exploit the wealth of knowledge they encapsulate.
It underscores the necessity for more effective approaches to task-specific analysis and demonstrates the effectiveness of our strategy of extracting knowledge from LLMs before reasoning directly to bridge these critical gaps in performance.
Furthermore, our findings suggest that GPT4o-based enhancement outperforms the other two LLMs on most datasets, indicating the potential of leveraging stronger LLMs to significantly enhance the effectiveness and performance of smaller models in various tasks.}

\textcolor{black}{Comparing MEC with COT, MEC emerges as the superior performer. This superiority can be attributed to the collaborative essence of MEC, where multiple experts collaborate to improve the accuracy and comprehensiveness of the knowledge extracted from the LLM, effectively mitigating the problem of inaccurate and incomplete knowledge due to hallucinations in the single LLM. Moreover, the MEC framework showcases its prowess in generating comprehensive and accurate insights through its iterative feedback and refinement mechanism.}

\textcolor{black}{In conclusion, the superior performance of LUK in log analysis tasks underscores its effectiveness across various datasets. By leveraging expert knowledge extracted from LLMs, LUK not only excels in task execution but also demonstrates the viability of enhancing smaller pre-trained models with sophisticated domain insights. This approach not only optimizes the utilization of the rich knowledge encapsulated within LLMs but also introduces a fresh paradigm in the realm of log analysis, paving the way for more efficient and insightful methodologies in this domain. In addition, LUK's process of acquiring knowledge is automated and independent of experts, making it more practical.
}

\begin{table}[]
\centering
\color{black}
\caption{Results of generalization ability experiments (SOFTWARE SYSTEM LOGS).}
\label{tab: beyond_pretrain}
\resizebox{\columnwidth}{!}{%
\begin{tabular}{lcc}
\toprule[1.5pt]
\multirow{3}{*}{Methods} & \multicolumn{2}{c}{Anomaly Detection (Precision / Recall / F1) }                                       \\ \cmidrule{2-3} 
                         & \multicolumn{1}{c|}{\multirow{2}{*}{ThunderBird}} & \multirow{2}{*}{Spirit} \\
                         & \multicolumn{1}{c|}{}                             &                         \\ \midrule
CNN                      & \multicolumn{1}{c|}{94.36 / 99.97 / 97.09}        & 97.55 / 97.11 / 97.33   \\
BiLSTM                   & \multicolumn{1}{c|}{98.10 / 96.96 / 97.52}        & 98.78 / 95.61 / 97.17   \\
BERT                     & \multicolumn{1}{c|}{98.35 / 97.21 / 97.78}        & \textbf{99.98} / 98.74 / 99.36   \\
Biglog                   & \multicolumn{1}{c|}{97.04 / 96.33 / 96.68}        & 99.56 / 99.07 / 99.31   \\
KnowLog                  & \multicolumn{1}{c|}{99.56 / 94.46 / 96.95}        & 99.38 / 99.20 / 99.29   \\ \midrule
T5-KD                    & \multicolumn{1}{c|}{99.35 / 95.08 / 97.17}        & 98.91 / 99.63 / 99.27   \\ \midrule
LLama3-70B               & \multicolumn{1}{c|}{55.05 / 93.96 / 69.42}                            & 78.16 / 97.49 / 86.76                       \\
ChatGPT                  & \multicolumn{1}{c|}{48.21 / 93.10 / 63.52}                            & 79.19 / 98.20 / 87.68                       \\
GPT-4o                   & \multicolumn{1}{c|}{49.89 / \textbf{100.0} / 66.57}                            & 75.44 / 90.32 / 82.21                       \\ \midrule
LUK (COT-ChatGPT)        & \multicolumn{1}{c|}{98.55 / 95.06 / 96.77}        & 99.81 / 98.92 / 99.37   \\
LUK (COT-GPT4o)          & \multicolumn{1}{c|}{99.35 / 95.08 / 97.17}        & 98.92 / \textbf{99.81} / 99.37   \\ \midrule
LUK (MEC-LLama3)         & \multicolumn{1}{c|}{\textbf{99.78} / 96.72 / \textbf{98.23}}        & 99.09 / 99.81 / 99.45   \\
LUK (MEC-ChatGPT)        & \multicolumn{1}{c|}{99.57 / 96.51 / 98.02}        & 99.89 / 99.16 / 99.52   \\
LUK (MEC-GPT4o)          & \multicolumn{1}{c|}{99.57 / 96.92 / 98.23}        & 99.27 / \textbf{99.81} / \textbf{99.54}   \\ \bottomrule[1.5pt]
\end{tabular}%
}
\end{table}

\begin{table*}[]
\centering
\color{black}
\caption{Results of generalization ability experiments (network device logs).}
\label{tab: beyond_pretrain2}
\resizebox{\textwidth}{!}{%
\begin{tabular}{lcc|ccccc|c}
\toprule[1.5pt]
\multirow{3}{*}{Methods} & \multicolumn{2}{c|}{Module Classifiction} & \multicolumn{5}{c|}{Log and Description Semantic Matching}                                                              & LPCR                  \\ \cmidrule{2-9} 
                         & \multicolumn{2}{c|}{H3C}                  & \multicolumn{1}{c|}{Cisco}         & \multicolumn{1}{c|}{Huawei}        & \multicolumn{3}{c|}{H3C}                      & Huawei                \\
                         & Switches            & Routers             & \multicolumn{1}{c|}{Security}      & \multicolumn{1}{c|}{Security}      & Switches      & Routers       & Security      & Security              \\ \midrule
CNN                      & 69.49 / 67.55       & 70.72 / 69.71       & \multicolumn{1}{c|}{70.36 / 70.36} & \multicolumn{1}{c|}{81.73 / 81.73} & 83.29 / 83.19 & 83.60 / 83.59 & 82.30 / 82.30 & 56.05 / 79.07 / 69.95 \\
BiLSTM                   & 70.21 / 68.45       & 71.40 / 69.93       & \multicolumn{1}{c|}{74.01 / 74.01} & \multicolumn{1}{c|}{79.12 / 79.12} & 80.88 / 80.83 & 83.81 / 83.80 & 82.57 / 82.57 & 55.65 / 79.21 / 69.80 \\
BERT                     & 81.11 / 79.78       & 77.93 / 76.05       & \multicolumn{1}{c|}{79.08 / 79.08} & \multicolumn{1}{c|}{89.22 / 89.22} & 87.44 / 87.41 & 88.25 / 88.25 & 85.94 / 85.94 & 67.89 / 89.73 / 79.55 \\
Biglog                   & 81.59 / 79.53       & 79.50 / 77.80       & \multicolumn{1}{c|}{82.73 / 82.69} & \multicolumn{1}{c|}{94.19 / 94.19} & 89.40 / 89.37 & 91.11 / 91.11 & 92.02 / 92.02 & 75.78 / 92.23 / 84.59 \\
KnowLog                  & 81.60 / 80.27       & 79.50 / 79.22       & \multicolumn{1}{c|}{84.94 / 84.93} & \multicolumn{1}{c|}{96.45 / 96.45} & 93.78 / 93.78 & 91.85 / 91.85 & 93.65 / 93.65 & 83.94 / 91.71 / 89.77 \\ \midrule
T5-KD                    & 82.32 / 81.26       & 80.41 / 80.15       & \multicolumn{1}{c|}{84.63 / 84.63} & \multicolumn{1}{c|}{95.31 / 95.31}  & 92.86 / 92.86 & 94.50 / 94.49 & 93.78 / 93.78 & 84.86 / 91.57 / 90.19 \\ \midrule
LLama3-70B               & 79.90 / 78.35       & 79.28 / 78.22       & \multicolumn{1}{c|}{89.42 / 89.42} & \multicolumn{1}{c|}{96.98 / 96.98} & 94.58 / 94.57  & 96.08 / 96.08 & 95.67 / 95.67 & \textbf{96.97} / \textbf{98.94} / \textbf{98.11} \\
ChatGPT                  & 77.48 / 73.77       & 76.35 / 74.10       & \multicolumn{1}{c|}{85.42 / 85.38} & \multicolumn{1}{c|}{90.87 / 90.76} & 91.01 / 90.96 & 91.96 / 91.90 & 93.24 / 93.24 & 94.95 / 97.01 / 96.56 \\
GPT-4o                   & 79.90 / 78.35       & 79.95 / 79.20       & \multicolumn{1}{c|}{\textbf{93.43} / \textbf{93.42}} & \multicolumn{1}{c|}{\textbf{98.49} / \textbf{98.49}} & 96.43 / 96.43 & \textbf{97.56} / \textbf{97.56} & 96.49 / 96.49 & 96.44 / 98.28 / 97.76 \\ \midrule
LUK (COT-ChatGPT)        & 83.05 / 81.52       & 80.63 / 79.00       & \multicolumn{1}{c|}{84.31 / 84.30} & \multicolumn{1}{c|}{96.33 / 96.33} & 93.89 / 93.89 & 94.18 / 94.18 & 93.91 / 93.91 & 82.23 / 94.73 / 88.83 \\
LUK (COT-GPT4o)          & 83.05 / 82.18       & 80.41 / 82.29       & \multicolumn{1}{c|}{84.94 / 84.94} & \multicolumn{1}{c|}{96.52 / 96.51} & 94.70 / 94.70  & 94.81 / 94.81 & 93.92 / 93.92 & 83.81 / 90.78 / 89.48 \\ \midrule
LUK (MEC-LLama3)         & 84.50 / 83.33       & 83.33 / 81.71       & \multicolumn{1}{c|}{87.64 / 87.63} & \multicolumn{1}{c|}{97.39 / 97.39} & 95.97 / 95.97 & 96.61 / 96.61  & 96.22 / 96.21 & 87.50 / 94.07 / 92.34 \\
LUK (MEC-ChatGPT)        & 83.78 / 82.69       & 83.11 / 81.40       & \multicolumn{1}{c|}{88.90 / 88.90} & \multicolumn{1}{c|}{96.99 / 96.99} & 95.16 / 95.16 & 95.34 / 95.34 & 95.54 / 95.54 & 86.71 / 93.55 / 91.75 \\
LUK (MEC-GPT4o)          & \textbf{85.47} / \textbf{84.57}       & \textbf{84.01} / \textbf{82.29}       & \multicolumn{1}{c|}{89.06 / 89.06} & \multicolumn{1}{c|}{97.52 / 97.52} & \textbf{96.54} / \textbf{96.54} & 96.93 / 96.93 & \textbf{96.89} / \textbf{96.89} & 87.36 / 94.47 / 92.22 \\ \bottomrule[1.5pt]
\end{tabular}%
}
\end{table*}

\begin{itemize}
    \item \textbf{RQ2: How effective is LUK in generalization ability?}
\end{itemize}
As the system evolves, many previously unseen logs will be collected \cite{zhang2019robust}, 
\textcolor{black}{and it is unrealistic to collect all the real-world logs for pre-training. In this RQ, to verify the generalization of LUK, we conduct experiments on downstream tasks where the logs in these datasets do not appear in the pre-training phase. Specifically, for software logs, we collect ThunderBird and Spirit as the datasets of the Anomaly Detection task from Loghub \cite{zhu2023loghub}. For network device logs, as shown in Table \ref{tab:network_statistics}, we collect logs from other vendors (H3C) and devices (Security) as the datasets.}

The experimental results are shown in Table \ref{tab: beyond_pretrain}, it can be found that LUK still has superior performance on the pre-training unseen logs, 
\textcolor{black}{especially on the task of Module Classification, LUK improves 3.38\% on average Accuracy compared to T5-KD.
These results strongly suggest that the expert knowledge distilled from LLMs assumes a pivotal role in empowering smaller PLMs to develop a profound comprehension of logging principles and mechanisms. This proficiency ensures that the PLM does not succumb to overfitting tendencies during training, a pitfall that impedes other fine-tuned models from effectively analyzing new logs by failing to grasp crucial information essential for accurate log analysis. }

\textcolor{black}{
In comparison to KnowLog, LUK still has a clear advantage over KnowLog, which takes knowledge from documentation.
The expert knowledge derived from LLMs offers a deeper level of comprehension and learning, enabling the PLM to generalize more effectively to new log datasets. Conversely, KnowLog, constrained by information within documents, may not provide sufficiently profound domain insights to support robust generalization of the PLM to unseen data.}

\textcolor{black}{
In essence, our findings affirm that LUK excels in its generalization capabilities by adeptly analyzing previously unseen logs. The infusion of expert knowledge sourced from LLMs equips the smaller PLM with the ability to assimilate and leverage vital insights from expert knowledge, thereby averting overfitting and bolstering its performance in log analysis tasks. This underscores the pivotal role of expert knowledge from LLMs in improving the adaptability of LUK to navigate the complexities of log analysis with finesse and precision.
}

\begin{figure}[]
    \centering
    \vspace{-2mm}
    \subfloat[Anomaly Detection (BGL)]{\label{fig: system type forgetting}\includegraphics[width=0.95\linewidth]{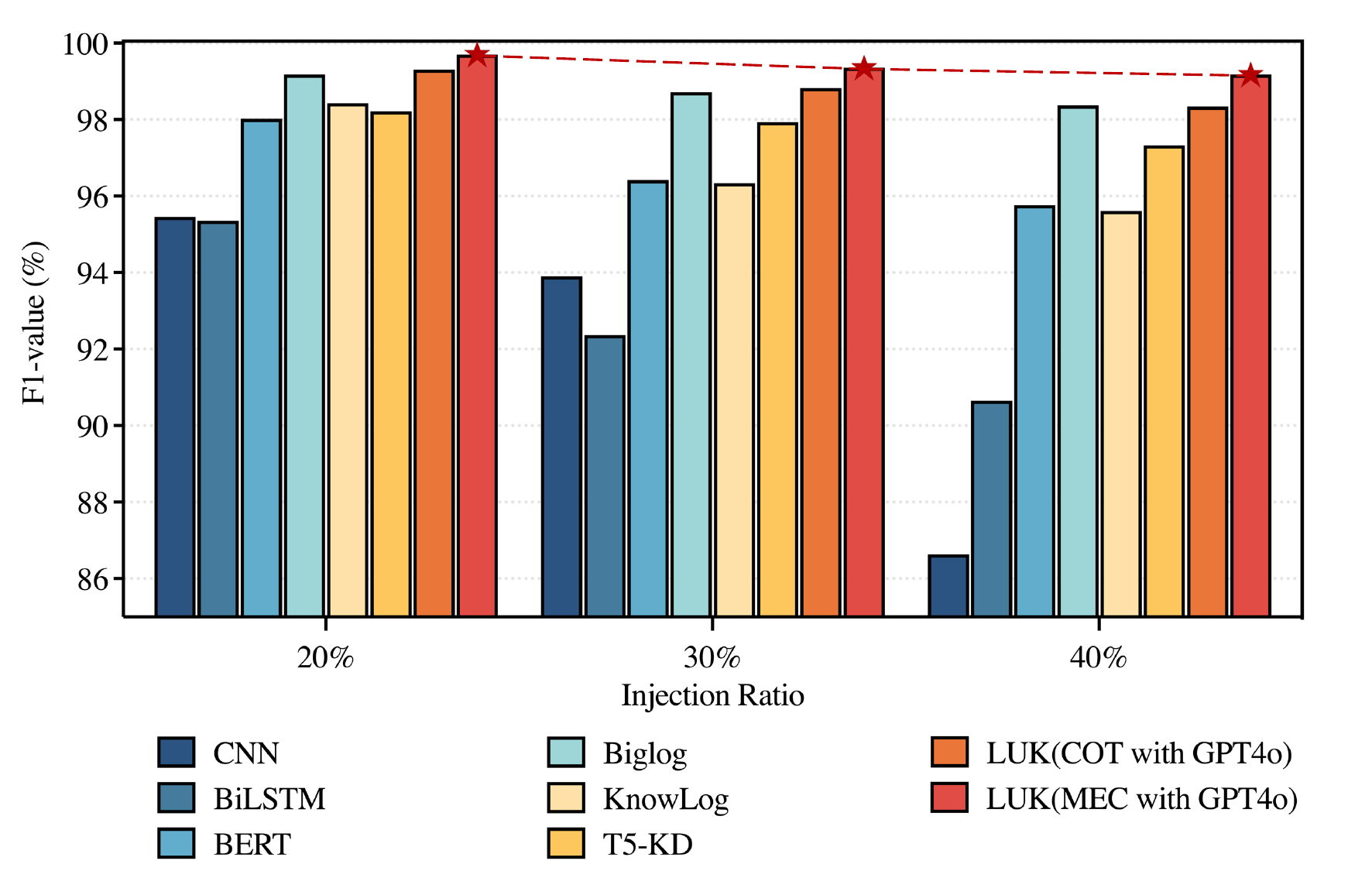}}
    
    \subfloat[LDSM (Huawei Switches)]{\label{fig: system type zero}\includegraphics[width=0.95\linewidth]{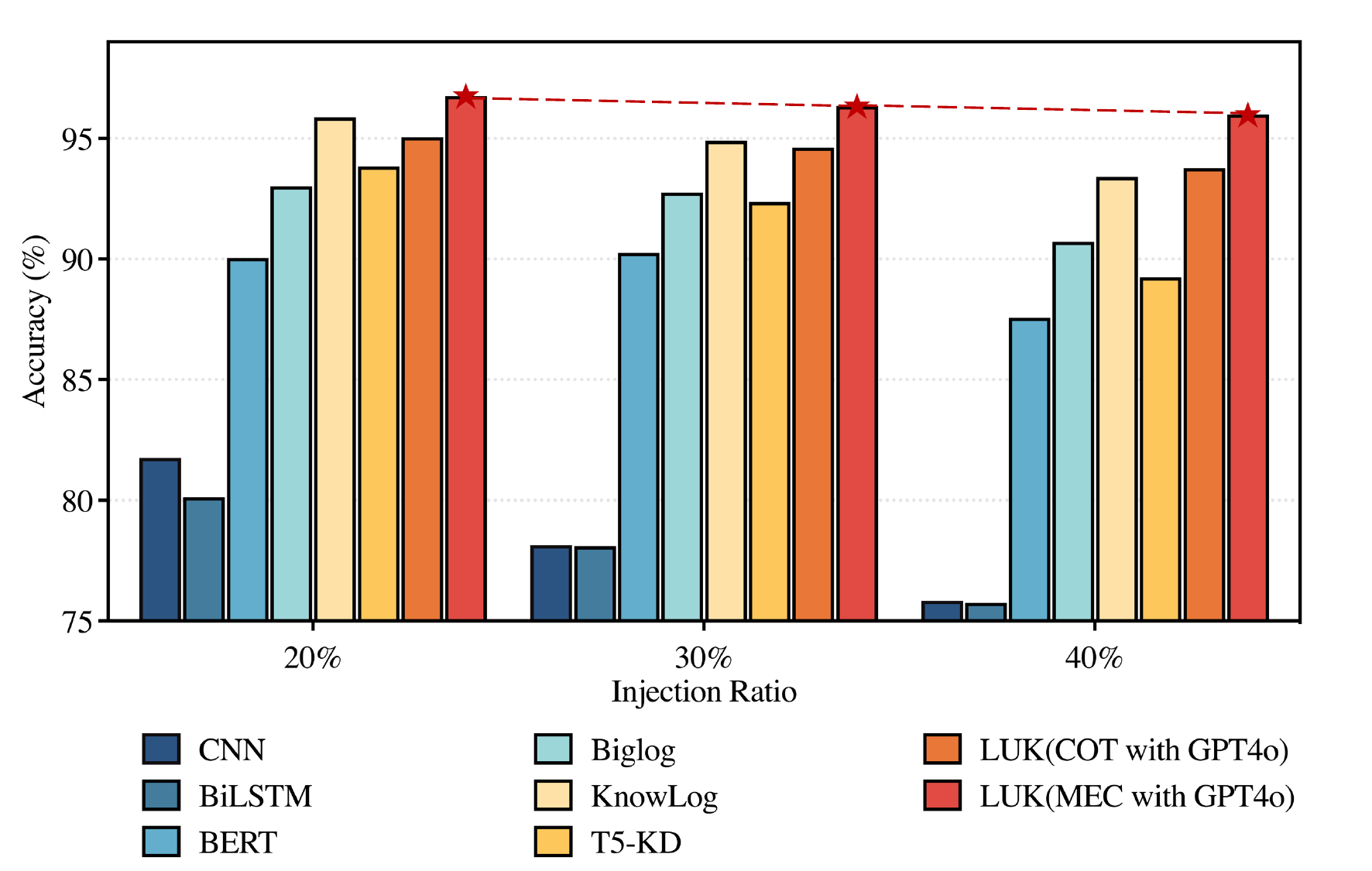}}
    \caption{\textcolor{black}{Results on different synthetic datasets of unstable log.}}
    \label{fig:robust}
\end{figure}

\begin{figure*}[]
    \centering
    \subfloat[Anomaly Detection (BGL)]{\label{fig: system type 
    forgetting}
    \includegraphics[width=0.33\linewidth]{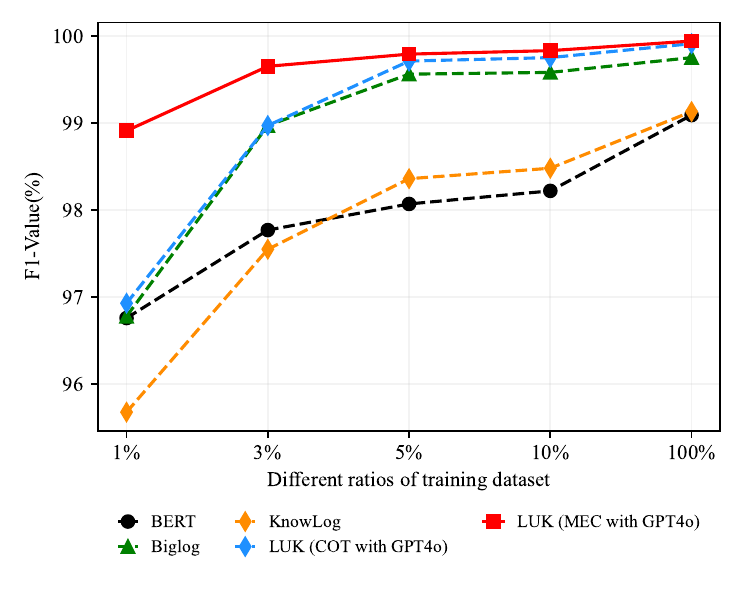}}
    \subfloat[LDSM (Huawei-Security)]{\label{fig: system type zero}
    \includegraphics[width=0.33\linewidth]{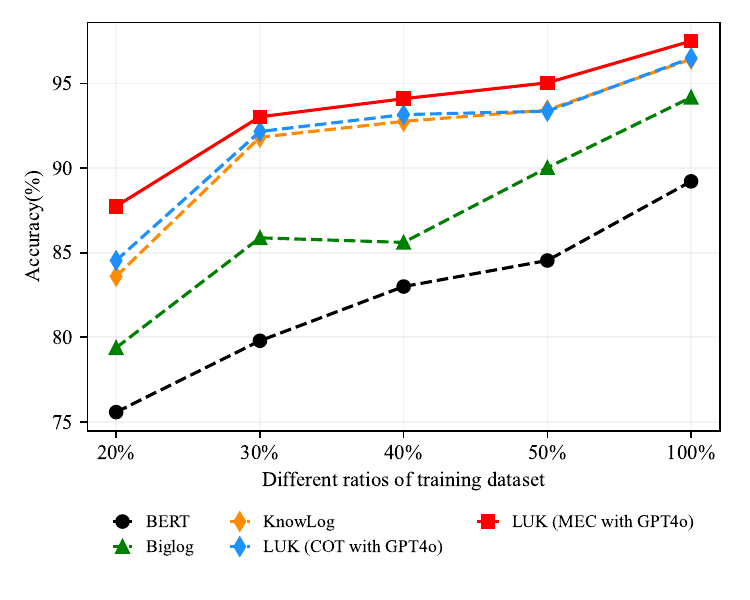}}
    \subfloat[Fault Phenomenon Identification]{\label{fig: system type zero}
    \includegraphics[width=0.33\linewidth]{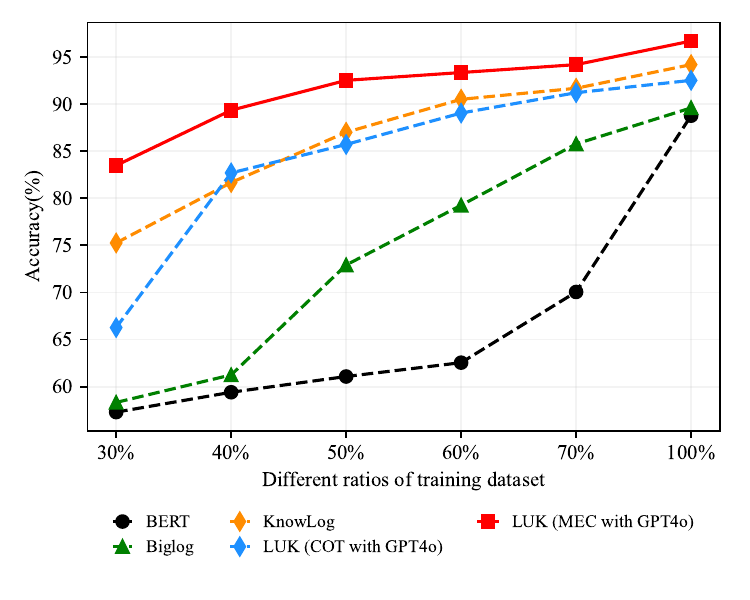}}
    \caption{\textcolor{black}{Results on different ratios of the training dataset.}}
    
    \label{fig:low-resource-results}
\end{figure*}

\begin{itemize}
    \item \textbf{RQ3: How effective is LUK on unstable log data?}
\end{itemize}
In real-world systems logs are \textit{unstable}, meaning that new but similar logs often appear, this is caused by the fact that developers may frequently modify the logging statements in source code. According to the investigation \cite{kabinna2018examining}, around 20\% - 45\% of logging statements may change throughout the lifetime. 
In this RQ, to evaluate the effectiveness of LUK on unstable logs, we conduct experiments on two log analysis tasks with unstable logs.  Following \cite{zhang2019robust}, we create two synthetic datasets to reflect the unstable characteristics of real-world logs, which are based on the BGL dataset of Anomaly Detection and Huawei-Switches dataset of Log and Description Semantic Matching. Specifically, we simulate unstable logs by randomly inserting or removing a few random words in the original log with different injection ratios.

The experimental results are shown in Fig. \ref{fig:robust}, it can be seen that LUK performs much better than other baselines. With the increasing injection ratio of unstable logs, the performance of all methods has declined in different degrees. However, MEC-based LUK declines relatively smoothly and still maintains high performance even under a high injection ratio. 
\textcolor{black}{
Specifically, as the injection ratio increased from 20\% to 40\%, LUK's F1 and Accuracy decreased by merely 0.54\% and 0.78\% on the anomaly detection and LDSM tasks, respectively. It confirms that LUK is robust enough to the unstable logs. The reason is that incorporating expert knowledge from LLMs into the smaller model assists in noise filtering, thereby enhancing the accuracy of log analysis for the smaller pre-trained model.}
Compared with CNN and BiLSTM, traditional methods perform the worst on unstable logs, which suggests that the limited semantic understanding of traditional methods hinders their ability to analyze logs more robustly. 
Compared with BERT and Biglog, although pre-training on log corpus can further improve log understanding, models are still limited to understanding logs with professional knowledge. 
\textcolor{black}{
In contrast, KnowLog's reliance on knowledge solely extracted from the documentation may lead to limitations in handling the evolving nature of log data. MEC-based LUK with its multi-expert collaboration, outperforms KnowLog by providing more deep and adaptable expert knowledge for robust log analysis tasks.
When compared to the COT-based methods, including T5-KD and COT-based LUK, MEC-based LUK showcases superior robustness on unstable logs. The collaborative essence of MEC, by incorporating multiple expert perspectives and iterative feedback loops, MEC ensures the refinement and validation of expert knowledge for log analysis tasks, enhancing robustness and accuracy.}

\textcolor{black}{Consequently, we can conclude that  LUK showcases remarkable robustness when confronted with unstable logs, demonstrating that leveraging expert knowledge from LLMs equips smaller PLMs with the necessary insights to navigate the challenge of robust log analysis, thereby enhancing their resilience against noise and instability. This highlights the pivotal role of expert knowledge in fortifying the log analysis capabilities of models operating in dynamic and evolving log environments.}

\begin{itemize}
    \item \textbf{RQ4: How effective is LUK with limited labeled logs?}
\end{itemize}
In real-world scenarios, acquiring a considerable quantity of annotated samples is difficult \cite{le2022log,le2023log}, thereby presenting a significant obstacle to the efficacy of automated log analysis models. To assess the effectiveness of LUK in low-resource tasks, characterized by limited annotations,
\textcolor{black}{we conduct experiments on Anomaly Detection, Log and Description Semantic Matching, and Fault Phenomenon Identification with different ratios of training datasets.}

\textcolor{black}{The experimental results are shown in Fig. \ref{fig:low-resource-results}, considering that knowledge distillation based on fewer samples are difficult to train, we do not compare with T5-KD.}
We find that with the reduction of training samples, the performance of various models exhibits a decline, \textcolor{black}{with BERT and Biglog demonstrating a particularly significant downward trend in the LDSM and FPI tasks, and KnowLog showing a significant decrease on the AD task}. On the other hand, LUK based on MEC achieves optimal results by fine-tuning the models with different proportions of annotated samples. 
\textcolor{black}{Specifically, on the AD task, compared to the full data, LUK drops only 1.03\% in the F1-value with 1\% of the training data. And on the FPI task, LUK's Accuracy drops by only 13.2\% with 30\% of the training data, while BERT and Biglog drop by 31.45\% and 31.25\%, respectively.
This demonstrates that LUK gains expert knowledge from LLM, which helps to compensate for the shortcomings of the smaller pre-trained model when less annotated data is available. Models with restricted knowledge struggle to excel in scenarios with limited resources, highlighting the significance of leveraging knowledge to mitigate the need for extensive annotation and enhance performance in data-scarce environments.
KnowLog primarily relies on knowledge extracted from documentation, which may be limited in scope and quality, especially in low-resource settings. In contrast, this proves that knowledge acquired by the LLM has a wider scope and can provide a more comprehensive and diverse range of information, rather than being limited to some specific documentation.}
Compared with utilizing COT to acquire knowledge, on the FPI task with 30\% of the training data, the Accuracy of MEC-based LUK  is 17.2\% higher than COT. This suggests that the MEC framework is more effective, which can be inferred that more rational and accurate knowledge enables the model to gain an advantage in low-resource scenarios. 

It is worth noting that the FPI task is a real scenario dataset, which cannot be analyzed directly with LLMs considering privacy issues. By using log templates to obtain expert knowledge from LLMs, a task-specific model is constructed based on LUK, which proves the effectiveness of LUK and improves the efficiency of task analysis.


\textcolor{black}{In summary, LUK achieves outstanding performance in low-resource scenarios, which stems from its ability to leverage expert knowledge acquired from LLMs, effectively compensating for the constraints of smaller PLMs when annotated data availability is restricted. By strategically harnessing this expert knowledge, LUK enables the development of efficient and accurate models even in resource-constrained environments, showcasing its adaptability and effectiveness in challenging log analysis tasks. }

\begin{figure}[]
    \centering
    \subfloat[LDSM (Huawei-Switches)]{\label{fig: system type zero}
    \includegraphics[width=0.95\linewidth]{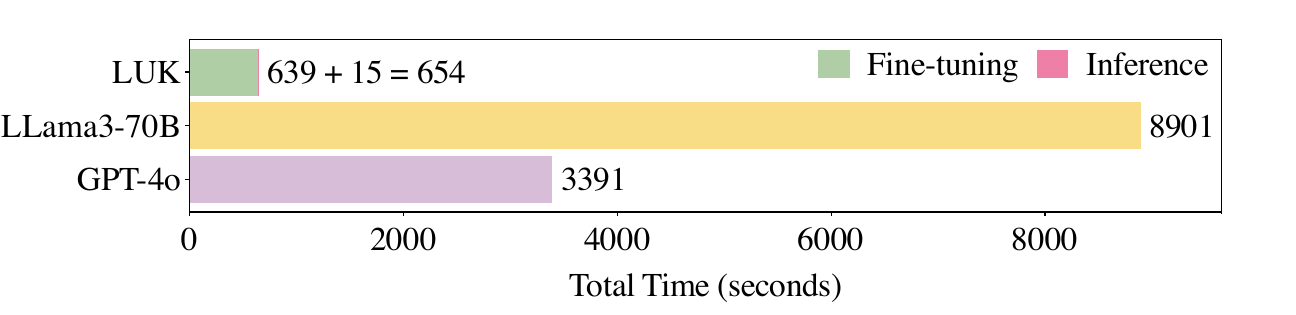}}
    
    \subfloat[LPCR (Huawei-Switches)]{\label{fig: system type 
    forgetting}
    \includegraphics[width=0.95\linewidth]{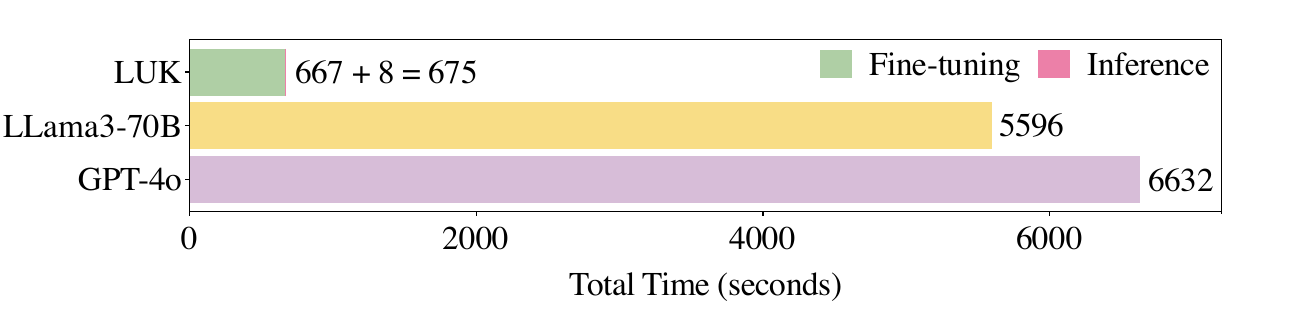}}

    \subfloat[Anomaly Detection (BGL)]{\label{fig: system type zero}
    \includegraphics[width=0.95\linewidth]{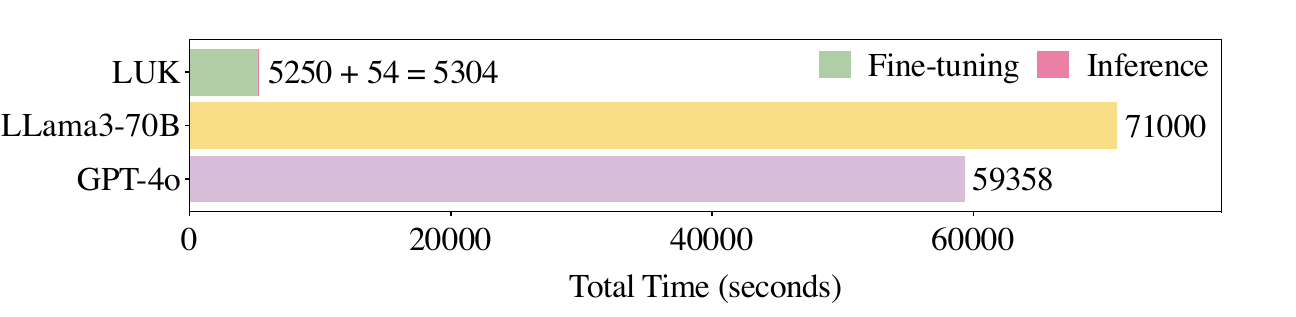}}
    
    \caption{\textcolor{black}{Time expense of different models during reasoning.}}
    
    \label{fig:time-cost}
\end{figure}

\begin{itemize}
    \item \textbf{RQ5: How efficient is LUK in inference compared with LLMs?}
\end{itemize}

Efficiency plays a crucial role in log analysis for practical applications, considering the vast volumes of logs \cite{le2023log}. In this RQ, we compare LUK's deployment and time costs with LLMs in inference. 
\textcolor{black}{
Specifically, we perform experiments on Anomaly Detection, Log and Description Semantic Matching, and Log and Possible Cause Ranking three tasks to calculate the inference time with all testing samples.  For fairness, we deploy LLama3-70B and LUK on the same hardware setup, consisting of 2 NVIDIA A100 GPUs, with a batch size of 1 for each model. It's worth noting that deploying LUK requires about 2 GB of GPU-memory, while deploying LLama3-70B requires 150 GB of GPU-memory.  For GPT-4o,  we utilize
HTTP requests to invoke the APIs then calculate time. 
}
Additionally, we leverage vLLM\footnote{\url{https://github.com/vllm-project/vllm}} to accelerate the inference process on LLama3.

\textcolor{black}{The results are shown in Fig. \ref{fig:time-cost}, we notice that 
LUK is significantly faster than GPT-4o and LLama3. Specifically, excluding the fine-tuning time, the average inference speed of LUK is 718x, and 868x faster than GPT-4o and LLama3-70B, respectively. 
Even considering the fine-tuning time, the average inference speed of LUK is still 9x, and 11x  faster than GPT-4o and LLama3-70B.
In addition, the average response time for each input is 4.13, and 4.69 seconds for GPT-4o and LLama3-70B respectively, while LUK's average response time is 0.0058 seconds. This swift response time positions LUK as a more practical solution for real-world log analysis scenarios. This efficiency can be attributed to the fact that LLMs possess extensive parameter sizes and generate outputs sequentially, necessitating more computational steps to deliver results. In contrast, LUK, with its smaller parameter sizes and more professional log analysis capabilities, can swiftly address specific tasks following knowledge enhancement and fine-tuning.}

\textcolor{black}{In conclusion, LLMs demand higher deployment resources and exhibit slower inference speeds. On the other hand, the knowledge-enhanced LUK offers rapid responses to inputs with limited deployment resources, making it a highly recommended choice for executing specific tasks in real-world scenarios. This advantage underscores the practicality and efficiency of LUK in comparison to traditional LLMs like GPT-4o and LLama3-70B.}

\begin{figure*}[]
    \centering
    \subfloat[Anomaly Detection (BGL)]{\label{fig: roundaaa}
    \includegraphics[width=0.33\linewidth]{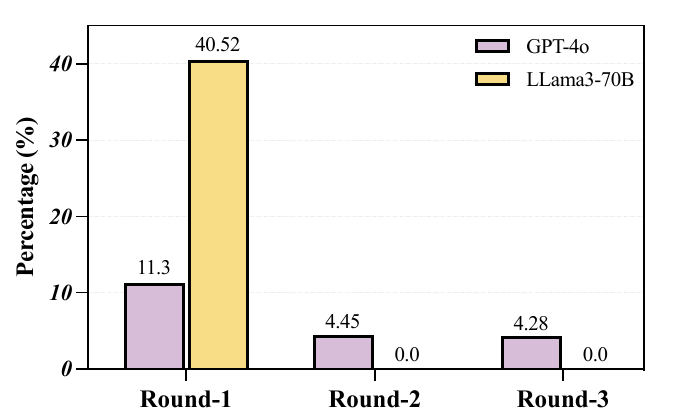}}
    \subfloat[MC (Huawei-Switches)]{\label{fig: system type zero}
    \includegraphics[width=0.33\linewidth]{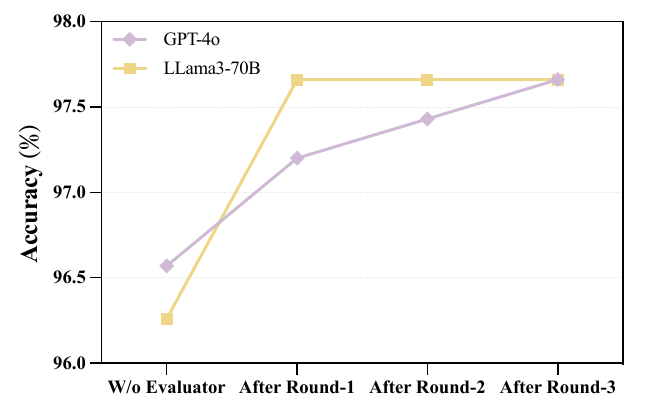}}
    \subfloat[LDSM (Huawei-Switches)]{\label{fig: system type zero}
    \includegraphics[width=0.33\linewidth]{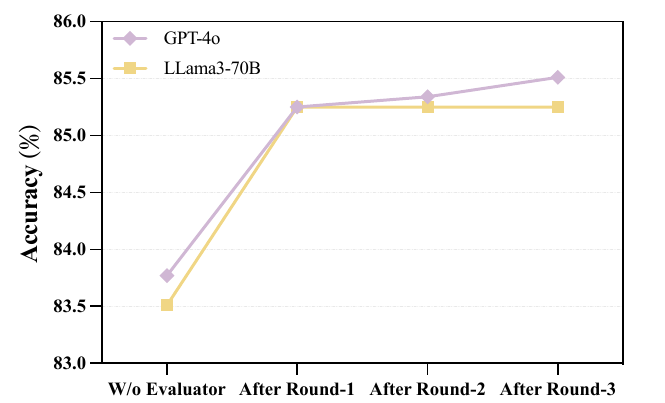}}
    \caption{\textcolor{black}{Figure (a) shows the proportion of unsatisfactory knowledge in each iteration. Figure (b) and (c) show the results of the refined knowledge based on the evaluation feedback each time.}}
    
    \label{fig:roundn}
\end{figure*}

\begin{table}[]
\centering
\color{black}
\caption{\textcolor{black}{Ablation studies for LUK, where TP and SA represent the pre-training tasks Token Prediction, and Semantic Alignment for LUK, respectively. MLM represents the pre-training task Masked Language Modeling of BERT and AP represents the pre-training task Abbreviation Prediction of KnowLog.}}
\label{tab:ablation}
\resizebox{\columnwidth}{!}{%
\begin{tabular}{lcc|cc}
\toprule[1.5pt]
\multirow{3}{*}{Methods}         & \multicolumn{2}{c|}{MC (Accuracy / Weighted-F1)} & \multicolumn{2}{c}{LDSM (Accuracy / Weighted-F1)} \\ \cmidrule{2-5} 
                                 & \multicolumn{2}{c|}{Huawei}                      & \multicolumn{2}{c}{Huawei}                        \\
                                 & Switches                & Routers                & Switches                & Routers                 \\ \midrule
BERT                             & 82.37 / 81.20           & 81.18 / 79.20          & 93.18 / 93.18           & 90.06 / 90.05           \\
LUK (MEC-GPT4o)                  & \textbf{85.51} / \textbf{85.26}           & \textbf{85.33} / \textbf{84.73}          & \textbf{97.66} / \textbf{97.66}           & \textbf{97.34} / \textbf{97.34}           \\ \midrule
$\vdash$ w/o Evaluator           & 83.77 / 83.26           & 82.30 / 81.08          & 96.57 / 96.57           & 95.53 / 95.53           \\
$\vdash$ w/o Evaluation Examples & 84.38 / 84.17           & 82.13 / 81.06          & 96.64 / 96.64           & 95.63 / 95.63           \\ \midrule
$\vdash$ w/o TP                  & 83.25 / 82.65           & 81.66 / 79.92          & 95.95 / 95.95           & 95.68 / 95.68           \\
$\vdash$ w/o SA                  & 83.77 / 83.19           & 84.14 / 82.92          & 95.25 / 95.25           & 94.68 / 94.68           \\ \midrule
$\vdash$ only with MLM           & 83.24 / 83.25           & 82.36 / 81.19          & 94.19 / 94.19           & 93.62 / 93.61           \\
$\vdash$ only with AP            & 83.60 / 82.89           & 82.72 / 81.68          & 94.39 / 94.39           & 93.97 / 93.97           \\ \bottomrule[1.5pt]
\end{tabular}%
}
\end{table}

\subsection{Ablation Studies}
To verify the effectiveness of the multi-expert collaboration framework and knowledge-enhanced pre-training tasks in LUK, we perform ablation experiments on MC and LDSM, which are two typical tasks of multi-class classification task and semantic matching task. The results are shown in Table \ref{tab:ablation}, where we notice that: Overall, LUK achieves optimal performance with the complete modules. The absence of any module can lead to performance degradation, which proves that MEC and pre-training tasks contribute positively. 

\textcolor{black}{Further, in our multi-expert collaborative framework, the \textit{Evaluator} plays a crucial role in evaluating the content generated by the \textit{Executor} to ensure the completeness and accuracy of the output expert knowledge. We remove the \textit{Evaluator} role and directly utilize the content generated by the \textit{Executor}. The results indicated a performance decrease in the absence of the \textit{Evaluator}, emphasizing the pivotal role of feedback from the \textit{Evaluator} in producing high-quality knowledge and reducing errors. 
In addition,  to verify the effectiveness of the contrastive examples for the \textit{Evaluator},  we conduct an ablation analysis by removing the referenced examples for the \textit{Evaluator} and requiring evaluations to be conducted in the reference-free setting. 
The experimental results reveal a substantial decline in the performance of the LUK  when lacking referenced examples, even surpassing the negative impact observed when the \textit{Evaluator} role was entirely removed.
This outcome suggests that the lack of referenced examples deprives the evaluation process of essential guidance and benchmarks, increasing the quality evaluation challenges faced by the system.}

\textcolor{black}{To verify the performance of the iterative evaluation and refinement mechanism, we first statistics the proportion of unsatisfactory knowledge in each iteration, and then verify the performance of the refined knowledge based on the evaluation feedback each time on the MC and LDSM tasks, where we set up 3 iterations in our experiment.
As shown in Fig. \ref{fig:roundn}, we can find a consistent decline in the proportion of unsatisfactory knowledge with each successive iteration. Simultaneously, the performance of the enhanced model exhibits steady improvement, underscoring the positive impact of the iterative evaluation and refinement mechanism within the framework. Notably, our observations indicate that a significant portion of knowledge refinement occurs following the initial feedback round, with subsequent iterations primarily focusing on refining a smaller fraction of the knowledge. To save costs, we suggest that the framework can iterate one time as a cost-effective strategy,  effectively balancing performance gains with resource optimization.}

For pre-training tasks, the performance of the model declines without either token prediction (TP) or semantic alignment (SA) pre-training task, which suggests that pre-training on logs with knowledge enhancement helps the model capture knowledge from external information and improves log understanding. Specifically, without the TP task, the model drops more significantly on the MC task, which implies that understanding domain tokens can enhance log understanding. And lacking the SA task, the model drops more remarkably on the LDSM task, which implies that external knowledge can enrich the contextual information of logs.
\textcolor{black}{
To further verify our proposed word-level Token Prediction (TP) task, we also compare it with other word-level pre-training tasks, including BERT's Masked Language Modeling (MLM) task and KnowLog's Abbreviation Prediction (AP) task. The results demonstrate the superior capability of our token prediction task in integrating expert knowledge effectively. 
Unlike MLM and Abbreviation Prediction tasks, our proposed pre-training task requires models to not only predict tokens within the log itself but also dynamically assimilate contextual insights from domain-specific knowledge using the knowledge perception module.
This distinctive aspect sets our approach apart, as BERT solely relies on internal information and cannot leverage external knowledge. Conversely, KnowLog is limited in its scope, focusing solely on abbreviations and failing to comprehend other critical log elements such as parameters, conditions, and more, which are essential for a comprehensive understanding of domain-specific terminology.}

\section{Discussions}

\begin{table*}[]
\centering
\caption{Qualitative examples of LUK and baselines. We calculate the cosine similarity as the score to analysis.}
\label{tab:example}
\tabcolsep=0.07cm
\resizebox{0.9\textwidth}{!}{%
\begin{tabular}{l|c|l|ll}
\toprule[1.5pt]
Label                    & Index              & \multicolumn{1}{c|}{Examples}                                                                                                & Models & Score          \\ \midrule
\multirow{6}{*}{Match}   & \multirow{3}{*}{1} & \textbf{Log:} \textcolor{black}{OPSA/3/OPS\_}CLI\_CLOSE\textcolor{black}{\_FAIL: Failed to} stop the terminal \textcolor{black}{using the script.}                                             & BERT   & 0.671          \\
                         &                    & (Script={[}script-name{]}, event={[}event-name{]}, instance={[}instance-id{]},terminal={[}cli-id{]})                         & Biglog & 0.624          \\
                         &                    & \textbf{NL Description:} There was a failure in stopping the terminal using a specific script.                                        & LUK    & \textbf{\color{black}0.740} \\ \cmidrule{2-5} 
                         & \multirow{3}{*}{2} & \textbf{Log:}  SYSTEM/4/SYS\_IMAGE\_ERROR: The next startup image \textcolor{black}{package is error}. (imageIndex={[}imageIndex{]},                                                  & BERT   & 0.723          \\
                         &                    &  curImageName={[}curImageName{]}, nextImageName={[}nextImageName{]}, errReason={[}errReason{]}) & Biglog & 0.629          \\
                         &                    & \textbf{NL Description:} An error occurred in the next startup image package.                                                         & LUK    & \textbf{\color{black}0.788}          \\ \midrule
\multirow{6}{*}{UnMatch} & \multirow{3}{*}{1} & \textbf{Log:} \textcolor{black}{OPSA/3/OPS\_}TERMINAL\_WRITE\textcolor{black}{\_FAIL: Failed to} display the string on the                                                  & BERT   & 0.657          \\
                         &                    & terminal \textcolor{black}{using the script.} (Script=xx,event=xx, instance=xx, string=xxx, terminal=xx)                                        & Biglog & 0.646          \\
                         &                    & \textbf{NL Description:} There was a failure in stopping the terminal using a specific script.                                        & LUK    & \textbf{\color{black}0.514} \\ \cmidrule{2-5} 
                         & \multirow{3}{*}{2} & \textbf{Log:} TRILL/4/TRILL\_RECV\_ERR\_PKT: TRILL-INFO: Drop \textcolor{black}{error packet.} (PktType={[}PktType{]},                                   & BERT   & 0.671          \\
                         &                    & ProcessId={[}ProcessId{]}, ErrReason={[}ErrReason{]}, ErrCount={[}ErrCount{]}, InterfaceName={[}InterfaceName{]})            & Biglog & 0.486          \\
                         &                    & \textbf{NL Description:} An error occurred in the next startup image package.                                                                         & LUK    & \textbf{\color{black}0.077}          \\ \bottomrule[1.5pt]
\end{tabular}%
}
\end{table*}

\subsection{Qualitative Analysis}
To show the usefulness of LUK more intuitively,  we select two representative cases for qualitative analysis. Specifically, we employ a pre-trained model to obtain embedding representations of the input log and natural language (NL) description, then calculate cosine similarity as the score to demonstrate their capability of semantic understanding. As shown in Table \ref{tab:example}, to provide a challenge in the case, we deliberately select logs with closer events and then compute their similarity to the description, which is matched only with one log.

From this case, we notice that: For the matched example, BERT demonstrates better performance, whereas Biglog's performance is comparatively weaker when compared to BERT. This suggests that only pre-training on logs makes it challenging to capture log semantics adequately without external knowledge and this may enlarge the semantic differences between logs and natural language. In contrast, LUK achieves the highest similarity score on the matched example and the lowest similarity score on the unmatched example, which can infer that benefiting from expert knowledge, LUK exhibits superior proficiency in capturing log semantics for log understanding. It also verifies the effectiveness of knowledge acquisition from LLMs. For the unmatched example, both BERT and Biglog exhibit significantly weaker performance compared to LUK, which implies that BERT and Biglog have limitations in recognizing log semantics. These results reveal the remarkable performance of LUK in log understanding.

\begin{figure*}[]
\centering
   \includegraphics[width=\textwidth]{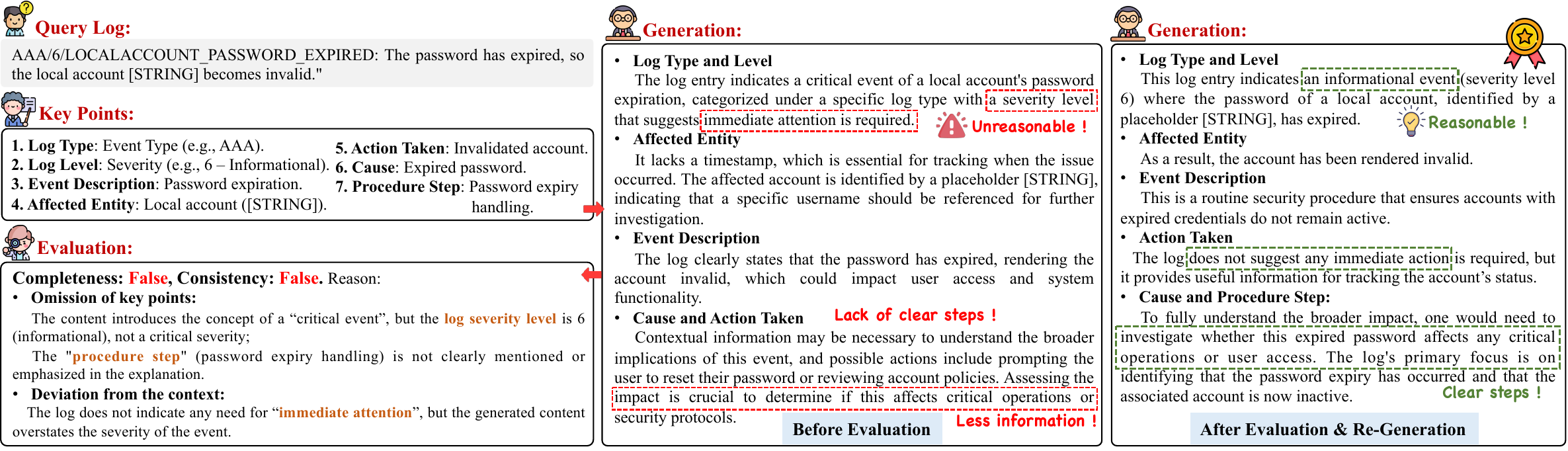}
   \caption{\textcolor{black}{An complete workflow of the MEC framework to construct knowledge.}}
   \label{fig:evaluation_case}
\end{figure*}

\begin{figure}[]
    \centering
    \subfloat[GPT-4o]{\label{fig: system type 
    forgetting}
    \includegraphics[width=0.5\linewidth]{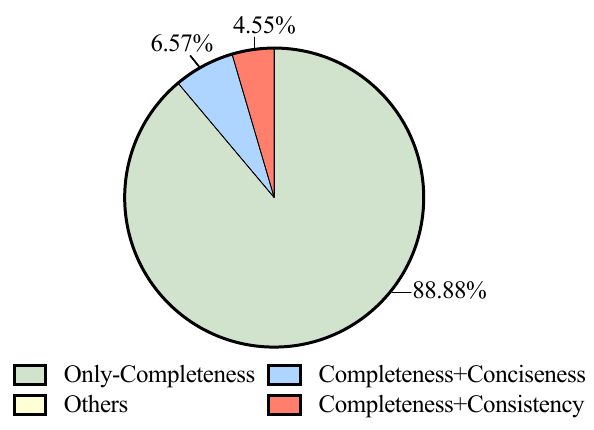}}    
    \subfloat[LLama3-70B]{\label{fig: system type zero}
    \includegraphics[width=0.5\linewidth]{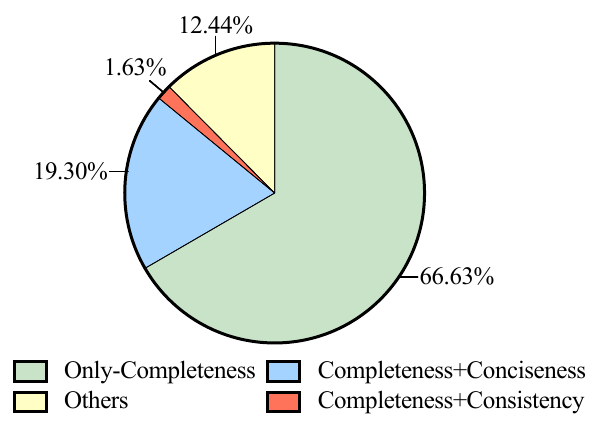}}
    
    \caption{\textcolor{black}{Distributions of error categories in unsatisfactory knowledge.}}
    
    \label{fig:different_error_precent}
\end{figure}

\subsection{Case Analysis}
\subsubsection{\textcolor{black}{Evaluation and refinement mechanism of the  MEC framework}}
\textcolor{black}{To intuitively understand the workflow of the MEC framework and analyze the impact of the evaluation and refinement mechanism on knowledge generation, we provide a specific case of expert knowledge construction as shown in Fig. \ref{fig:evaluation_case}, this input log from the Huawei device\footnote{\url{https://support.huawei.com/enterprise/en/doc/EDOC1100332302/fb54bda3}}.}

\textcolor{black}{
As can be seen from the case, the \textit{Director} first meticulously analyzes the input log to give key points, and then the \textit{Executor} generates expert knowledge based on the guidance of these key points. Following knowledge creation, the \textit{Evaluator} evaluates the quality of the generated knowledge and provides feedback to the \textit{Executor} for refinement. The evaluation feedback shows that the content generated by the \textit{Executor} does not fulfill completeness and consistency. Specifically, the generated knowledge lacks of an introduction to the procedure step and the explanation of the severity level is not rigorous. In addition, there is also an uninformative explanation in the generated knowledge, which makes it more difficult for the smaller model to understand the expert knowledge.
Finally, the \textit{Executor} refines the generated knowledge based on the evaluation feedback. From the re-generated knowledge, we can find that the \textit{Executor} not only supplements the clear introduction of the procedure step but also modifies the explanation of severity level to make it more reasonable. Overall, the re-generated knowledge is more comprehensive and rigorous,  which is more conducive for the smaller model to learn expert knowledge.}

\textcolor{black}{To more deeply reveal the issues of knowledge generated by LLMs. We count the distributions of different error categories in unsatisfactory knowledge according to \textit{Evaluator}'s results for GPT-4o and LLama3-70B, respectively. 
As shown in Fig. \ref{fig:different_error_precent}, it can be found that a significant portion of the inadequacies stems from issues related to completeness. The prevalence of completeness-related errors underscores the challenge faced by the single LLM in producing comprehensive knowledge autonomously. Moreover, alongside completeness issues, a smaller yet notable proportion of unsatisfactory knowledge is attributed to deficiencies in conciseness and consistency. This finding underscores the significance of collaborative frameworks that integrate diverse roles, each contributing unique perspectives and expertise to compensate for the capacity deficiencies of the single LLM and improve the completeness and accuracy of generated knowledge.}

\begin{figure*}[]
\centering
   \includegraphics[width=\textwidth]{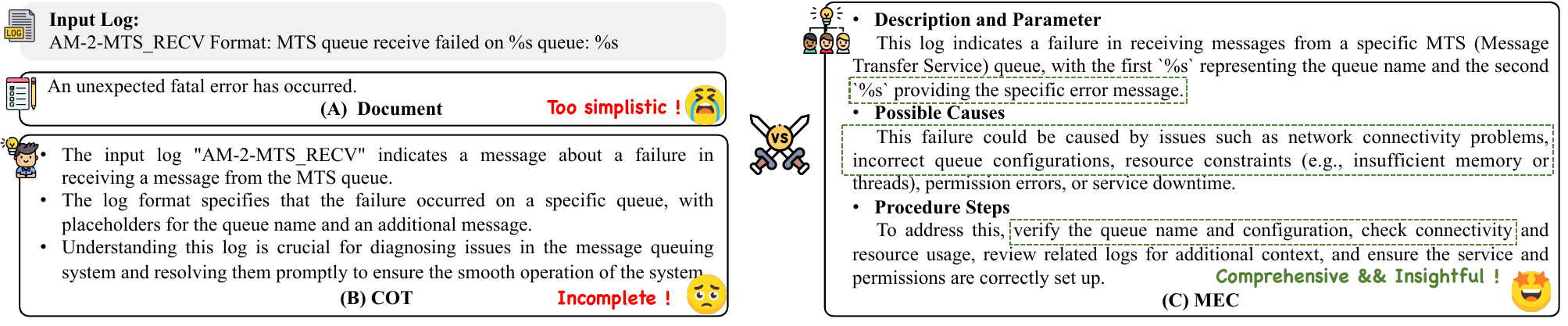}
   \caption{\color{black}Example of knowledge generated by GPT-4o and retrieved from documentation.}
   \label{fig:knowledge_case}
\end{figure*}

\begin{figure}[]
\centering
    \includegraphics[width=\columnwidth]{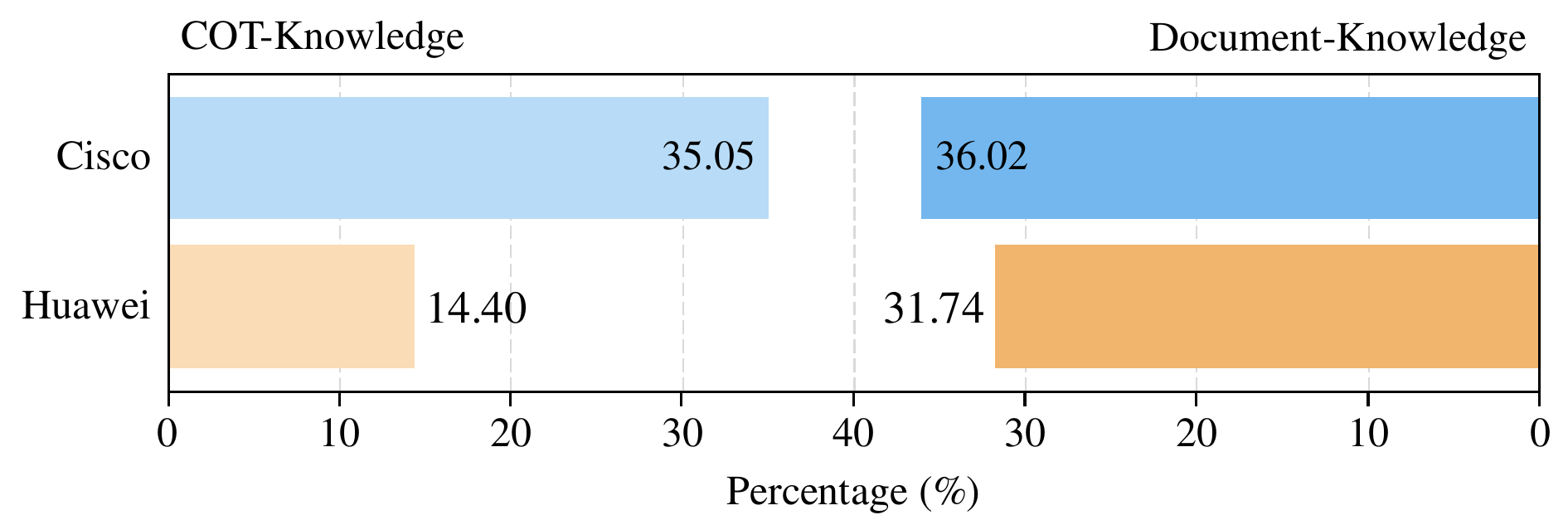}    
    
    \caption{\textcolor{black}{Proportion of unsatisfactory knowledge, where we employ the evaluation criteria of the MEC framework to evaluate the knowledge.}}
    
    \label{fig:doc-knowledge-analysis}
\end{figure}

\subsubsection{\textcolor{black}{Comparison of knowledge from different sources}} 
\textcolor{black}{To further intuitively analyze the quality of the expert knowledge constructed with the MEC framework, we compare it with the knowledge collected from documentation and the COT-based approach. As shown in Fig. \ref{fig:knowledge_case}, we collect a Cisco log\footnote{\url{https://www.cisco.com/c/en/us/td/docs/switches/datacenter/sw/routing_messages/reference/7k_rout_mess_ref_book/7k_rout_mess_ref_2mess.html}} as the case and give the corresponding generated knowledge based on the COT and MEC methods.}

\textcolor{black}{
As can be seen from these cases, the knowledge provided by the documentation is too simple, which provides a generic and limited understanding of the log without delving into specific details or potential causes. Although COT-based knowledge offers a more detailed explanation of the log, compared to MEC-based knowledge, there is still the problem of inadequate knowledge and lack of insight.
In contrast, the knowledge derived from the MEC framework excels in its comprehensive analysis and insights. By explicitly outlining the specifics of the log error, including the MTS queue failure and the placeholders for queue and error message, the MEC-based knowledge provides a deeper understanding of the issue at hand. Furthermore, it goes beyond mere description to identify potential root causes such as network connectivity issues.
The comparison highlights the clear advantage of the MEC framework in expert knowledge generation. Its collaborative nature allows for the integration of diverse expertise and perspectives, resulting in a more comprehensive analysis of complex logs.}

\textcolor{black}{
To quantify the differences among the various forms of knowledge, we employ the evaluation criteria of the MEC framework to evaluate the knowledge acquired by the documentation and the knowledge constructed based on COT in the network device logs. The evaluation results are shown in Fig. \ref{fig:doc-knowledge-analysis}, revealing a notable contrast in quality between the different sources of knowledge. It is apparent from the analysis that a significant proportion of knowledge sourced from documentation falls short of meeting the evaluation requirements. This deficiency can be attributed to the inherent limitations of documentation-derived knowledge, which often lacks depth and detail, offering only the most straightforward description of the logs. Combined with Fig. \ref{fig: roundaaa}, this suggests that the knowledge acquired from MEC based on LLMs is comparable or even better than the documentation. Considering the lack of documentation support for most logs and the high cost of manually constructing knowledge, this provides a feasible way to gain a wealth of practical knowledge automatically.}

\textcolor{black}{
Despite the advantages of the MEC framework in knowledge generation, domain-specific documentation continues to hold an indispensable role. LLMs remain struggling with knowledge gaps, especially when dealing with customized modules and terminology, leading to misconceptions. In such cases, domain documentation can bridge these gaps, providing essential context and expertise to enhance the understanding of LLMs. This symbiotic relationship allows LLMs to enrich documentation content and, in turn, enables documentation to improve the professionalism and accuracy of LLMs. We will optimize the integration of domain documentation and LLM to further enhance the knowledge acquisition process in the future.}



\subsection{Threats to Validity}
\subsubsection{Construct Validity}
Since LLMs are black box models, LLMs pose a risk of generating output content that may be unreasonable, potentially impacting the accurate interpretation of logs.
\textcolor{black}{First, to counteract the inherent randomness in the inference process of LLMs, we set the model temperature to 0. This step ensures that LLMs consistently produce results for the same inputs, mitigating variability in output.}
Second, to mitigate hallucinations of LLMs, we propose a multi-expert collaboration framework, which can work better with the power of cooperation and interaction. In particular, we design an \textit{Evaluator} in this framework, which checks the generated knowledge by evaluating completeness, consistency, and conciseness. 
\textcolor{black}{In addition, to address the issue of ambiguous and inconsistent evaluations by LLMs under the reference-free setting, we construct contrastive examples consisting of positive and negative knowledge of the same log. By presenting these contrastive examples to the \textit{Evaluator} as reference examples, we steer the evaluation of LLMs toward our desired objectives. In cases where issues are identified, the \textit{Executor} within the framework revises or refines the generated content based on the feedback received. This iterative process helps improve the quality and reliability of the generated knowledge. We also verify the importance of the \textit{Evaluator} for acquiring high-quality knowledge in ablation studies.}


\subsubsection{Internal Validity}
It is widely agreed that the performance of DL models is significantly affected by hyperparameters. Due to the limited computation resources, we do not search for the optimal hyperparameter settings, and instead follow the empirical settings. We acknowledge that further fine-tuning of these hyperparameters may yield better results. 


\subsubsection{External Validity}
From the perspective of enterprises, utilizing external LLMs to acquire expert knowledge may cause leakage of user privacy and internal information in logs. To alleviate this issue, on the one hand, we propose to remove sensitive information from logs and utilize log templates after log parsing to acquire knowledge. On the other hand, LUK is a general framework that can combine any LLMs, and users can also employ their own LLMs to acquire knowledge to enhance a small model for solving a specific problem. 
\textcolor{black}{For example, apart from invoking the proprietary LLMs, we also verify the effectiveness of LUK by deploying LLama3-70B in our experiments, which is an open-source LLM.}

\section{Conclusion}
In conclusion, this paper introduces LUK, a novel knowledge enhancement framework to improve log understanding with expert knowledge from LLMs. Unlike existing LLM-based log analysis studies that directly use the in-context learning (ICL) of LLMs, LUK first acquires expert knowledge from LLMs, then enhances the log pre-training with the corresponding expert knowledge on a smaller pre-trained language model, finally the enhanced pre-trained model for logs can be fine-tuned to solve downstream log analysis tasks. 
Compared to existing models, LUK achieves state-of-the-art performance on different log analysis tasks, which proves that expert knowledge from LLMs can be used more effectively to understand logs.

\section{DATA AVAILABILITY STATEMENT}
Our source code and detailed experimental data are available at \url{https://github.com/LeaperOvO/LUK}.

\ifCLASSOPTIONcaptionsoff
  \newpage
\fi

\bibliographystyle{IEEEtran}
\bibliography{ref}

\end{sloppypar}
\end{document}